\documentclass[twocolumn]{aastex631}
\hypersetup{linkcolor=red,citecolor=blue,filecolor=cyan,urlcolor=magenta}

\usepackage[T1]{fontenc}
\usepackage{graphicx}
\usepackage{amsmath}
\usepackage{float}

\usepackage{nicematrix} 

\BeforeBegin{NiceTabular}{\let\begin\BeginEnvironment\let\end\EndEnvironment}
\BeforeBegin{NiceArray}{\let\begin\BeginEnvironment}
\BeforeBegin{NiceMatrix}{\let\begin\BeginEnvironment}

\shorttitle{Effect of accretion disk on GW-Emission}
\shortauthors{Chatterjee et al.}
\graphicspath{{./}{Graphics/}}

\begin{document}

\title{Ideal E/IMRI vs Real E/IMRI system : Observable signature in LISA}

\correspondingauthor{SC, SM, PB}
\email{sangita6chatterjee@gmail.com,soumenjuphysics@gmail.com, prasadcsp@gmail.com}

\author[0000-0003-3979-113X]{Sangita Chatterjee}
\affiliation{Department of Physics, Jadavpur University, Kolkata 700032, West Bengal, India}

\author{Soumen Mondal}
\affiliation{Department of Physics, Jadavpur University, Kolkata 700032, West Bengal, India}

\author{Prasad Basu}
\affiliation{Department of Physics, Cotton University, Guwahati, Assam, India.}

\begin{abstract}
Real extreme/intermediate mass ratio inspiral(E/IMRI) systems are likely to contain large accretion disks which could be as massive as the central supermassive black hole. Therefore, contrary to its ideal model, a real E/IMRI system contains a third important component: the accretion disk. We study the influence of these disks on the emitted GW profile and its detectability through proposed LISA observation. We use a semi-relativistic formalism in the Kerr background \citep{PhysRevD.73.064037,2008PhRvD..77j4027B} for the case of transonic accretion flow which is a potential candidate to describe the accretion flows around AGN. The hydrodynamic drag of the disks modified the motion of the companion as a result the emitted wave changes in amplitude and phase. We found that these changes are detectable through the last few years of observation by LISA (in some cases as small as six months) for EMRIs residing within 3 GPc from the detector and for the accretion rate of the primary black hole of the order of $\dot{M}=1 \dot{M}_{Edd}$. These choices of parameter values are consistent with real systems.  The drag effect and hence the detectability of the emitted GW is sensitive to the hydrodynamical model of the disk. Therefore such observations will help one to identify the nature of the accretion flow and verify various paradigms of accretion physics.
\end{abstract}


\section{Introduction}
\label{sec:intro}
Extreme or intermediate mass ratio inspirals (E/IMRI) are the main class of sources that the space-based detectors LISA \citep{2017arXiv170200786A} and Taiji \citep{2020IJMPA..3550075R} aim to detect. The amplitude and the frequency of the emitted gravitational wave from such systems are within the detector sensitivity range \citep{2021NatRP...3..344B}. An E/IMRI is a type of system where a comparatively smaller mass compact object (white dwarf, neutron star, black hole(BH)) orbits or inspiral around a supermassive black hole (SMBH). Thus the companion moves essentially as a test particle in the background space-time of the central SMBH. Therefore the emitted GW signal from such a system provides a map of the background metric of the central supermassive object. This opens up the possibility to explore physics in the near horizon region of the supermassive object enabling one to test the validity of the various paradigms in a strong gravity regime like existence of naked singularity, exotic stars, prediction of alternative theories of gravity, deformed BH spacetime, etc.

However, to achieve success in this ambitious endeavor both the accurate modeling of the companion's motion and the detection of GW waveform with extreme precision are required. Detection of GW relies on the matched filtering technique which requires a bank of theoretically produced templates, which in turn requires very precise modeling of the companion's motion around the central compact object. To reproduce theoretically the trajectory of the moving companion, one needs to know about all the interactions present in the system-the interaction with the central SMBH as well as the interaction with the environment. Thus one must include environmental effects to properly model the motion of the companion around central SMBH.

On the other hand, to extract information from the emitted GW signal one requires to detect and analyze the GW waveform with extreme precision. Since the dominant effects are all given by the general relativity and conventional theories of stellar structure, any signature of alternative theories of gravity or exotic stars will be visible as a correction to the leading order effects and thus will be extremely small \citep{2018GReGr..50...46B,2019IJMPD..2842002C}. Therefore, it is highly possible that environmental effects even if small compared to the leading order effects will be strong enough to be visible in and to modify the higher order effects significantly. Hence a proper detection of environmental effects is essential to do high-precision science using gravitational waves.  

In this work, we are focusing on an E/IMRI system and the real example of an E/IMRI system is the active galactic nuclei (AGN). An AGN residing in a gas-riched environment naturally accrete matter in the form of a disk known as accretion disk. These disk could be as massive as the central black hole(BH) itself \citep{2013CQGra..30x4008M}. Accretion disk is a source of observed EM radiation from the AGN. However the observability of EM radiation from gas near SMBHs depends on the wavelength of radiation and the source of interest. In addition to this EM-spectrum of the disk, GW can also provide an efficient and independent way to probe the inner region of the disk.

Therefore, instead of two point-like objects orbiting around each other in vacuum, a realistic E/IMRI is a system in which the lighter mass compact object moves around the central SMBH through the accretion disk surrounding the central SMBH. For such a system the main environmental effect arises from the hydrodynamic drag on the moving companion by the disk matter. The accretion flow presents here in the vicinity of AGN then is most likely to be one with very high accretion rate and low angular momentum. Also at a large distance from the central SMBH, the matter is supposed to be originally subsonic, terminally supersonic and at the event horizon, it becomes comparable to light velocity $c$ (note that maximum sound speed in relativistic regime is $\frac{1}{\sqrt{3}}c$ ). 
One of the very interesting classes of solutions of Euler equation for describing the above mentioned accretion around AGN is the transonic solutions with multiple sonic points.
Not only because it is derived in a mathematically consistent way using relativistic hydrodynamic equation with boundary conditions that are highly realistic but also because the formation of shock in the flow, produces a hot Comptonzing region naturally in the inner part of the disk which reprocesses the soft photons coming from the disk into the hard photons, describes non-thermal component of the specta very well unlike the other models where the region has been introduced in an ad hoc manner. Several studies earlier \citep{1981ApJ...246..314A,1987PASJ...39..309F,1989ApJ...347..365C,1990ttaf.book.....C,1996MNRAS.283..325C,1997A&A...321..665L,2010ApJ...708.1442M}) have used these steady-state solutions topologies to describe several observational features (like the existence of hard photons, QPOs, and jets, etc.). Recently in numerical simulations, an upgraded time-dependent version of this model has been made (TCAF) which are describing successfully the observational features of black hole binaries as well as of the AGN spectra \citep{2019ApJ...877...65N,10.1093/mnras/stab1699,2021Galax...9...21M,2022A&A...662A..77M,2022mgm..conf..231M,2022JApA...43...90M,2022A&A...663A.178M}

The hydrodynamic drag on the moving companion arises from two sources: the accretion drag, produced due to the exchange of energy and angular momentum of the companion through the accretion of matter from the disk and the other is the dynamical friction caused due to the density perturbation generated by the moving companion in the disk matter. Hydrodynamical drag due to accretion becomes particularly important when the angular momentum distribution of the disk flow is sufficiently non-Keplerian.

The preliminary studies of disk effects were mainly done in the Newtonian framework or in order of magnitude estimation and we see a clear significance of the chosen hydrodynamical model of accretion disk on the strength of drag effects from all the previous studies \citep{1993ApJ...411..610C,2000ApJ...536..663N}. The first complete relativistic treatment considering an accretion tori is done by \citet{2008PhRvD..77j4027B}. Soon after \citet{PhysRevLett.107.171103,2011PhRvD..84b4032K} also studied the accretion effects but by considering a Shakura-Sunyaev disk model \citep{1973A&A....24..337S}. The similarity in these two disk models is that both possess small or negligible radial velocity and moderate accretion rate  (sub-Eddington accretion). Interestingly an observable effect for LISA has been found first by \citet{2011PhRvD..84b4032K}. Very recent simulation studies by \citet{2019MNRAS.486.2754D} also found some observable effects in terms of phase shift. All these studies \citet{2019MNRAS.486.2754D,2022MNRAS.517.1339G} are considered different disk models which are relevant in certain contexts however some model exhibit few unrealistic features.

In this context, we extend the study of \citet{2008PhRvD..77j4027B} and did a semi-relativistic treatment in the case of transonic accretion disk which represents the thin disk model very well and also as mentioned above has very interesting features than other flow models. Therefore it will be very interesting to extend the study with transonic accretion flow to describe the disk’s hydrodynamics and see whether these non-Keplerian accreting flows exhibit more prominent drag effects and due to the presence of one or more sonic points along the flow, sudden changes of the drag force in the shock region will produce observable effects that reflect on the emitted GW and observable by LISA.

The rest of the paper is planned accordingly: in section \S \ref{prev}, a brief review on the effect of the disk drag effects is given, and in section \S \ref{gbh}, we discuss the model equations. In section \S \ref{tans}, we discuss the transonic disk solutions around central SMBH. In \S \ref{result}, we discuss the various results of our study for different elliptical orbits. We estimated dephasing of GW emission and calculate the SNR. Finally, we end up in Section \S \ref{dissc} with discussion and conclusion.

\section{Previous Study}
\label{prev}

The effect of a massive accretion disk on the emission mechanism and profile of gravitational waveform from E/IMRI is a long-standing issue. The study on this issue was initiated by the work of S.K. Chakrabarti in 1993 \citep{1993ApJ...411..610C,1996PhRvD..53.2901C,1994ApJ...436..249M} where they considered the effect of hydrodynamic drag due to the accretion of matter by the companion moving in a Keplerian orbit from a non-Keplerian disk. The study shows that in general the drag force reduces the infall time for sub-Keplerian disks and for super Keplerian disks the infall time increases and in some extreme cases can create a stationary orbit. Soon after Narayan \citep{2000ApJ...536..663N} did an order-of-magnitude study for the ADAF model and did not find any significant effect. However, the disk models of Chakrabarti and Narayan are distinctly different. More development in this field has been done by \citet{1993astro.ph..5034G,1993MNRAS.265..365V,1998MNRAS.298...53V,1999A&A...352..452S,2001A&A...376..686K,2007MNRAS.374..515L,2008MNRAS.388..219B,10.1007/978-3-319-94607-8_4} considering several different physical scenarios like on and off equatorial orbital motion, radiatively-efficient self-gravitating disk model or inefficient advection dominated accretion flow (ADAF) model, etc. All these studies have been done in the Newtonian or pseudo-Newtonian framework. The first study in the general relativistic framework was done by \citet{2008PhRvD..77j4027B} considering a thick torus with zero radial velocity. 

A similar but more detailed study considering a standard disk has been done by \citet{PhysRevLett.107.171103,2011PhRvD..84b4032K}. This work shows disk migration gives a significant influence on the E/IMRI. \citet{2014PhRvD..89j4059B,Barausse_2015} pointed out that for eLISA observation of E/IMRI, environmental disturbances are negligible except for geometrically thin disk. The result of all these studies shows that the effect of the accretion disk crucially depends on the choice of the hydrodynamic model of the accretion disk. Recent study using simulation by \citet{2019MNRAS.486.2754D,2019MNRAS.489.4860D,2021MNRAS.501.3540D}, where they have considered a $2-D$ viscous accretion disk in a Newtonian framework around an IMRI and found disk produce a phase shift which is detectable by LISA if the surface density is $10^3 gcm^{-2}$. However, in their studies, they considered the hydrodynamical model that represents the accretion flows which are very unlikely to appear in realistic situations.

\section{Motion of the companion object}
\label{gbh}

As mentioned, we consider an E/IMRI system, where the central SMBH (mass $M$) determines the curvature of the background space-time in which the companion black hole of mass $m(m\ll M)$ moves like a test particle. The unperturbed motion of the companion object is the geodesic of the background spacetime whose metric is completely determined by the central SMBH. This is a reasonable approximation because the mass of the companion $m \ll M$. The companion as a test particle follows the geodesic of the central black hole which evolves secularly under radiation reaction. When the binind energy of the orbit is much much greater than the energy emission rate per cycle due to gravitational wave the evolution of orbit can be determined using adiabatic approximation \citep{2002phrvd..66d4002g,2002PhRvD..66f4005G}.

We choose the rest frame of the central SMBH and in the Boyer-Lindquist (B-L) coordinate ($r$, $\theta$, $\phi$, $t$) the metric of the Kerr space-time took the form,

\begin{equation}
\begin{split}
 dS^2 &= g_{\mu \nu}dx^{\mu}dx^{\nu}  \\
      &=(1-\frac{2r}{\Sigma})dt^2+(\frac{4ar \sin^2 \theta}{\Sigma})dtd \phi - \frac{\Sigma}{\Delta}dr^2- \Sigma d \theta^2  \\
      & -(r^2+a^2+ \frac{2ra^2 \sin^2 \theta}{\Sigma})\sin^2 \theta d \phi^2 \\
 \end{split}
 \label{kerr}
\end{equation}
Here $\Delta =r^2-2r+a^2$ and $\Sigma =r^2+a^2 \cos^2 \theta$, $a$ is the Kerr parameter. Here we use $M=G=c=1$ units.
The fiducial geodesic of the background space-time can be defined by two scalar constants viz, the specific energy  $E=u_a \bigl (\frac{\partial}{\partial t}\bigr )^a$, and specific angular momentum  $L=u_a (\frac{\partial}{\partial \phi}\bigr)^a$, where $\bigl (\frac{\partial}{\partial t}\bigr )^a$  and $(\frac{\partial}{\partial \phi}\bigr)^a$ are the two killing vectors corresponding to the stationarity and axisymmetry of the background space-time respectively. Equation of geodesic then takes the form \citep{2002PhRvD..66f4005G},

\begin{equation}
 r^4 (\frac{dr}{d \tau})^2= T^2- \Delta(r^2+(L-aE)^2)=V_r  \label{radial1}
\end{equation}

\begin{equation}
 r^2 \frac{d \phi}{d \tau}=-(aE-L)+\frac{aT}{\Delta}=V_{\phi} \label{phi}
\end{equation}

\begin{equation}
 r^2 \frac{d t}{d \tau}=-a(aE-L)+\frac{(r^2+a^2)T}{\Delta}=V_t    \label{tt}
\end{equation}

In our case we are considering equatorial geodesic therefore,
\begin{equation}
 \theta(\tau)=\pi/2
\end{equation}
where $T=E(r^2+a^2)-La$, $\Delta=r^2-2r+a^2$.

\subsection{Loss rates due to GW}
\label{gloss}
The emitted GW carries energy($E$) and angular momentum($L$) from the system, and due to this companion loses 4-momentum $P^{\mu}$, and as a result, the orbits of the companion shrinks gradually. The associated change in the four-momentum of the companion can be estimated as 
\begin{equation}
\left. \frac{dP^{t}}{d\tau} \right|_{GW} =\left. \frac{dE}{d \tau}  \right|_{GW}= \left.\frac{dE}{dt} \right|_{GW}.u^t_{sat}  \nonumber
\end{equation}

\begin{equation}
\left. \frac{dP^{\phi}}{d\tau} \right|_{GW} = \left. \frac{dL}{d \tau} \right|_{GW}= \left. \frac{dL}{dt}  \right|_{GW}.u^t_{sat} \label{ELGW1}
\end{equation}

In the weak field approximations, the energy and angular momentum loss rates due to the emission of GW are given in \citet{1996PhRvD..53.3064R}.

\subsection{Energy and angular momentum loss in the presence of disk}
\label{diskEL}
The presence of accretion disk in E/IMRIs directly influences the loss rates. The effect of this could be understood by considering two situations : (a) the E/IMRI system without an accretion disk and (b) the E/IMRI system with the accretion disk. Comparing the results would help us to estimate the effect of the disk on the companion motion of E/IMRIs. In the first case, loss of energy and angular momentum of the companion occurs due to the emitted GW radiation, as a result, the orbital radius decreases, and coalescence takes place in a longer time. In case (b) the presence of the disk offers additional forces exerted on the companion which results in more energy and angular momentum loss compared to case (a). The higher loss rates help to shrink the orbital radius more rapidly and coalescence takes place comparatively faster. The additional forces that act on the companion are due to the companion's (i) self-accretion and (ii) dynamical friction which changes the 4-momentum of the companion.

The total momentum change suffers due to disk and GW emission is
\begin{equation}
 \left. \frac{dP^{\mu}}{d\tau} \right|_{Total}=\left. \frac{dP^{\mu}}{d\tau} \right|_{GW} + \left. \frac{dP^{\mu}}{d\tau} \right|_{Accr} + \left. \frac{dP^{\mu}}{d\tau} \right|_{Defl}
 \label{pmu}
\end{equation}

The first term is already discussed in the previous section \S \ref{gloss}.  The other components can be obtained from \citep{2007ApJ...665..432K,2007MNRAS.382..826B,2008PhRvD..77j4027B}.
Accumulating all the effects, the modified 4-momentum of the companion can therefore be obtained as,
\begin{equation}
 P^{\mu}=P^{\mu}+\left.\frac{dP^{\mu}}{d \tau}\right|_{total}d\tau
 \label{pmu1}
\end{equation}
where $d\tau$ is the proper time interval between two successive orbits of the companion.

\subsection{Phase shift acquired by the orbiting companion}
\label{phs}

The ambient medium leaves an imprint due to hydrodynamic drag on the trajectory of the companion. As a result, the infall time, orbiting frequency gets modified. This changes although very small leaves an accumulated effects in the total phase of the GW signal. This accumulated phase is thus different when the companion is not embedded within the disk. Thus, in the presence of the disk, one can calculate the difference in phases and use the result to identify the presence of a massive accretion disk around the central supermassive black hole. At any time the orbital frequency of the moving companion is $\omega$ is given by,

\begin{equation}
\omega=\dot{\phi}=\frac{d \phi}{dt} \label{dotphi}
\end{equation}
The frequency of the emitted gravitational wave is, therefore, $\omega_{GW}= 2 \omega= 2 \dot{\phi}$. Therefore the total phase accumulated during the decrease in orbital radius from $r_0$ to $r_1$, can be obtained by,
\begin{equation}
\phi = \int_{\tau_0(r_0)}^{\tau(r_1)}\omega dt
\end{equation}
Therefore the phase difference in the presence and in the absence of the disk would be,
\begin{equation}
\delta \phi= \phi_{\text{with-disk}} - \phi_{\text{no-disk}}
\end{equation}
 Thus the acquired phase can be obtained using the above formula and one can also identify the amount of phase shift of the companion.

\section{Accretion processes around central SMBH}
\label{tans}
The fluid dynamics of a non-self graviating disk is given by the energy momentum tensor of a viscous fluid. We ignored the self-gravity of the disk so that it does not change the curvature of the background spacetime. The stress-energy tensor is given by,
\begin{equation}
 T^{\mu \nu}=(\rho + p)u^{\mu}_{fluid}u^{\nu}_{fluid}+pg^{\mu \nu}+2\eta \sigma^{\mu \nu}
\end{equation}
where  $u_{fluid}^{\mu}=(u^t,u^r,u^{\phi},0)$  is the 4-velocity and $p$ is the pressure, and $\rho=\rho_0(1+ \mathcal{E})$ is the energy density of the fluid, $\rho_0$ and $\mathcal{E}$ are the rest-mass density and internal energy density per unit rest mass. $\eta$ is the coefficient of dynamical viscosity related to the kinematic viscosity $\nu$ by $\eta=\nu \rho_0$. $\sigma$ is given by,

\begin{equation}
 \sigma^{\mu \nu}=\frac{1}{2}\left(u^{\mu}_{;\beta}\mathrm{p}^{\beta \nu}+u^{\nu}_{;\beta}\mathrm{p}^{\beta \mu}\right)-\frac{1}{3}\Theta \mathrm{p}^{\mu \nu}
\end{equation}
$\Theta=u^{\mu}_{;\mu}$ and $\mathrm{p}^{\mu \nu}=u^{\mu}u^{\nu}+g^{\mu \nu}$ is the projection operator.

The hydrodynamics of the fluid is described by the Euler equation $ \nabla_{\mu} T^{\mu \nu} = 0$ and Continuity equation $(\rho_0u^{\mu}_{fluid})_{; \mu} = 0$. In the steady accretion flow, a perfect fluid obeys the two conserved equations: $ hu_{t}^{fluid}=E_{fluid}$ and $hu_{\phi}^{fluid}=-l_{fluid}$ where $h=\frac{p+\rho}{\rho_0}$ is the relativistic enthalpy.
Using these equations and integrating $\frac{du^r}{dr}$ and $\frac{da_s}{dr}$, we obtain the transonic flow solutions as described in \citet{1981ApJ...246..314A,1987PASJ...39..309F,1989ApJ...347..365C,1990ttaf.book.....C,1996MNRAS.283..325C,1996ApJ...471..237C,1997A&A...321..665L,2010ApJ...708.1442M} and references therein.

\begin{figure}
\centerline{\includegraphics[width=1.0\linewidth,clip]{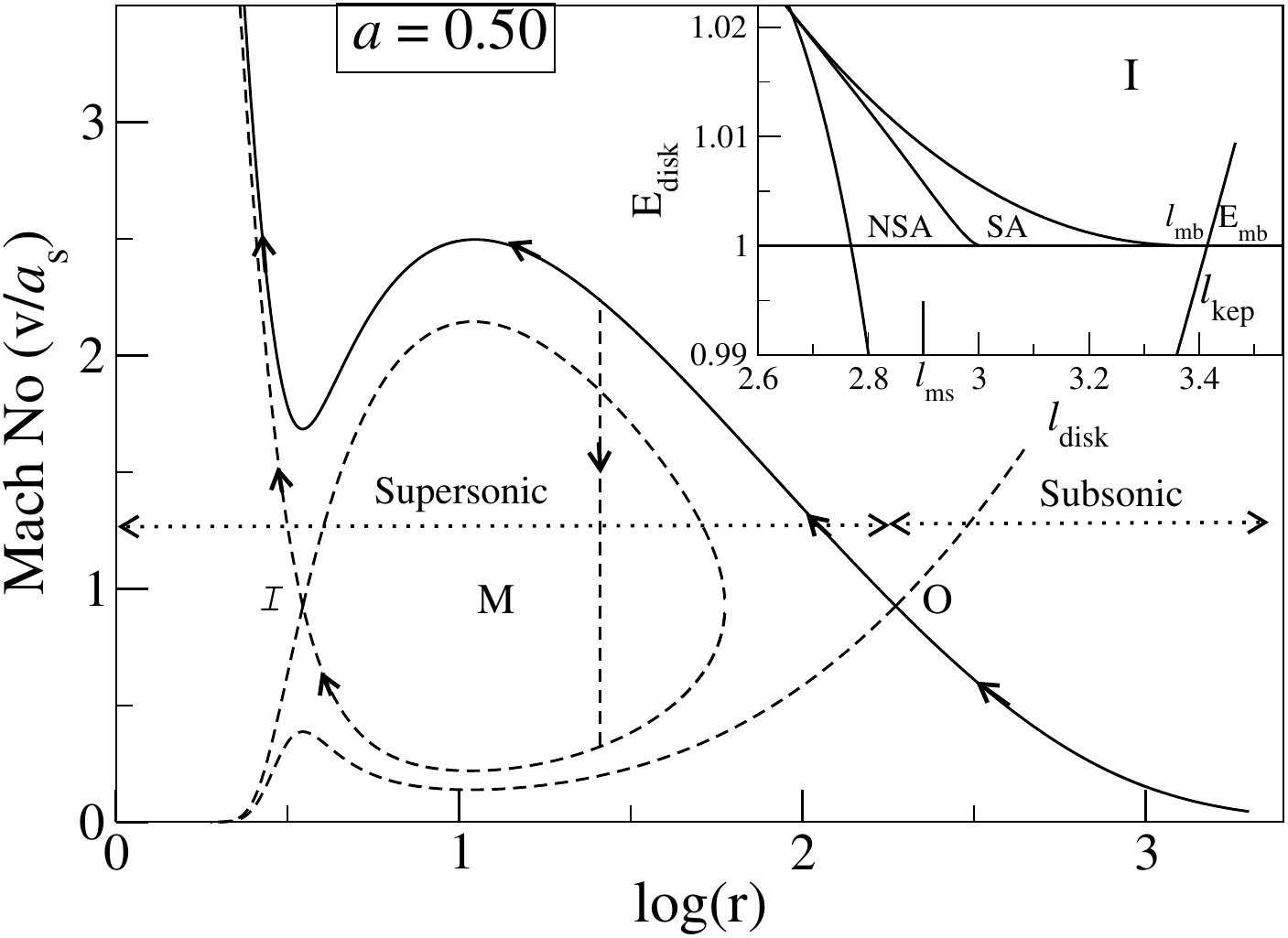}}
\caption{
\label{Soltran} { We plot the solution topology (Mach no. vs radial distance $r$)  with multiple sonic/critical points. The location of the three critical points are : $inner(\mathcal{I})=3.50r_g$(X-type), $middle(M)=10.77r_g$(O-type), $outer(O)=186.78r_g$(X-type). The solid arrow-headed line represents a transonic solution of the hydrodynamic equation. We see that at far distance, flow is subsonic, and becomes supersonic after crossing the outer critical point O. Flow may suffer a shock transition at $r_S=26.81r_g$ shown by the vertical arrow line. After the shock, flow becomes sub-sonic and then becomes supersonic again at the inner sonic point ($\mathcal{I}$) to satisfy the inner boundary condition at the BH horizon. The location of shock, determined by the Rankine-Hugoniot Condition, and that of the critical points depends on the specific energy($E_{disk}$), specific angular momentum ($l_{disk}$) of the flow. In the subpanel, we plot the parameter space of $E_{disk}$ \& $l_{disk}$ and its subdivisions w.r.t the number of sonic points. The NSA(no shock in accretion) and SA (shock in accretion) represent the parameters space of our main interest for which three sonic point exist. The point on the right curve $l_{kep}$ represents the solution for a Keplerian disk. The full parameter space, subdivision, and the respective solutions are discussed in detail in \citet{1996MNRAS.283..325C}. These transonic solutions are appropriate for AGN accretion because ambient matter at large distances is sub keplerian and subsonic. We choose $E_{disk}=1.003$, $l_{disk}=3.05< l_{kep}$, $a=0.50$  in our solution. }}
\end{figure}

We present here one such solution obtained from the shock in accretion region in figure \ref{Soltran}, where we plot Mach no. $\frac{v}{a_s}$ versus radial distances. The solid arrow headed curve is one of the transonic solutions which we have considered as a disk solution in the merger studies. We see that at large distance flow is subsonic, crosses the outer critical point($O$) at $r=186.78r_g$  becomes supersonic and then the supersonic flow reaches the black hole horizon\citep{1988PASJ...40..709K,1989PASJ...41..271N, 1984PASJ...36...71M}. However the accretion disk may harbour a shock\citep{1987PASJ...39..309F,1994MNRAS.270..871N, 1995A&A...295..238Y,Caditz_1998,1994PASJ...46..257N,1993ApJ...417..671C,1996ApJ...470..460M} in the flow before crossing the middle sonic point ($M$) at $r=10.77r_g$. The arrow-headed dotted line represents the shock in the solution in which the supersonic flow transit to subsonic branch and again becomes supersonic at the inner critical point $\mathcal{I}$ at $r=3.5r_g$\citep{2016ApJ...819..112L,2008ApJ...689..391N,1988PASJ...40..709K}. Important point is to note here that the solution satisfies correct boundary conditions $v=0 \to c$ as predicted in GTR and the flow has a substantial radial velocity between the outer critical point ($O$) to the horizon as shown in the figure, which is generally uncommon in other studies. This unique feature differs from the other solution. These solutions\citep{1995ApJ...455..623C,Chakrabarti:2016iln,Smith_2002} are therefore very important while studying the net drag force acted upon the companion during its merging.

\subsection{Observational evidences of transonic flows:}
In the early 1980s, these disks became favorite candidates for the explanation of the "big blue bump" seen in the UV region of the active galaxies (\cite{1982ApJ...254...22M,1989ApJ...346...68S} and references therein) with some success. With the advent of space-based observations, the hard and soft X-ray spectra of these objects show that there could be even bigger bumps in the X-ray regions of the spectra. It was becoming clear that the standard accretion disks, which are in general cooler, cannot explain the production of X-rays.

The explanations of the hard components seen in the spectra of black hole candidates required extra components, such as plasma clouds, hot corona, etc. which are self-consistently constructed and identified as a hotter post-shock region in transonic flows when flow passes through a shock transition\citep{1995ApJ...455..623C}. Hence, the  X-ray continuum emission or the non-thermal component of the power-law spectrum naturally produced from the hot post-shock corona of the accretion disk via a scattering process: (e.g., Seyfert I \& II, Quasar, Blazer, BL Lac). The power-law (PL) component is widely accepted as the effect of inverse Compton scattering of thermally produced soft photons from the accretion disk by a corona of hot electrons close to the inner edge of the disk. There are proposed models in the literature that introduced the corona in an ad-hoc manner, to be the region between the truncation radius and the ISCO radius. However, it is not clear why the disk truncates at a certain radius from the central black hole and how the truncation radius is connected with the corona. According to the transonic flows shock forms at the low angular momentum satisfying the Rankine-Hugoniot conditions, produces naturally a hot Comptonizing region by converting its substantial kinetic energy of the infalling matter to the thermal energy, and thus shock location is the radius of the truncation of the disk, often use to identify the inner boundary of the thin disk solution. A typical X-ray spectrum of AGN in the $2-10$ keV region shows primarily the signature of the formation of the hot Comptonizing region and provides an indirect evidence of transonic accretion flows. Specifically, an object such as M87 does not seem to have the big blue bump in its spectrum (which is a signature of the Keplerian disk), the spectral data of M87 may be satisfactorily well fitted by a sub-Keplerian component alone, and there is little need of any Keplerian component\citep{2008ApJ...689L..17M}.

Moreover, the same shock location, which can explain the spectral features, may also be able to explain the temporal features from its oscillation \citep{1996ApJ...457..805M}. The excess thermal pressure in the post-shock corona puffed up and pushes the matter to produce the outflows and high energy jets in AGN systems \citep{Le_2005,Chattopadhyay:2011rmr}

Furthermore, in an active galaxy, the incoming matter is either a mixture of Keplerian and sub-Keplerian flows, or entirely sub-Keplerian. This is because matter may be supplied from constantly colliding winds from a large number of stars $\sim$ pc away, or from a dusty torus at the outer edge and thus most of the angular momentum of the infalling matter may be lost unlike the case of accreting matter from its companion binary star where the matter supply is likely to be Keplerian. The transonic solutions, obtained in this context, well agree with the outer boundary of such objects. Numerical simulations \citep{2013MNRAS.430.2836G,10.1093/mnras/stv223} show that when the injected flow is sub-Keplerian, as is very likely in the case of AGNs, the enhanced viscosity on the equatorial plane, segregates the matter into two components(the equatorial Keplerian component and Sub-Keplerian halo outside) as envisaged by Chakrabarti in TCAF model \citep{1995ApJ...455..623C}. These physical properties and the dynamics of the flow make the transonic solution more favorable as a physical model than other existing models in the literature. Hence, this canonical thin-disk model is also used widely in the literature (e.g. for x-ray binaries) to extract physical parameters of the flow \citep{Chakrabarti:2016iln} and should be applicable for all the black hole candidates (from the usual quasars and AGNs to nano quasars or stellar-mass black holes)\citep{1995ApJ...455..623C,2008ApJ...689L..17M}. There
are also observational evidences that is relevant for SMBH accretion \citep{2019ApJ...877...65N,10.1093/mnras/stab1699,2021Galax...9...21M,2022A&A...662A..77M,2022mgm..conf..231M,2022JApA...43...90M,2022A&A...663A.178M}.

In spite of the above observational evidence, so far as we know, several models exist, are valid under certain assumptions, and are capable of describing a certain set of observational features. The current status of these models is that there is no single model which described all the observed features correctly. At the same time, there is not enough observational evidences to rule out any model completely. This is an open issue and unsettled question where no consensus is formed to date. We do not claim that one particular disk model prevails over other. The aim of our study is to see whether GW observation can shed some light on it or not, whether the effect of different hydrodynamic models can be manifested on the GW observation or not, whether it helps to solve this issue. Moreover, since the transonic flow that we considered in our study is one of the potential candidates to describe the accretion flow around SMBH. Therefore, study of disk effect from GW emission will remain incomplete if one does not include this transonic model.

From our finding we see that the phase difference is very sensitive to the selected model of hydrodynamic disk. One may constrain the disk parameters from this particular parameter.

\section{Results}
\label{result}

Here we consider the inspiral stage of a binary black hole where the mass ratio is $EMR = \frac{m}{M}$. To calculate the fluxes averaged over one period and energy and angular momentum loss rate $\frac{dE}{dt}$ and $\frac{dL}{dt}$ for the GW emission, we use the quadrupole approximation formula given by \citep{Ryan:1995zm}. The background metric of the system is given by a Kerr metric. We here assume the combined metric of the binary system is solely due to the central supermassive Kerr BH (Equation \eqref{kerr}) and is not affected by the presence of the companion and disk. The disk is geometrically thin in which the transonic accretion flow solution is considered around the central SMBH having accretion rate $\dot{m}=\frac{\dot{M}}{\dot{M}_E}$, where $\dot{M}$ is the accretion rate of the companion BH, $\dot{M}_E$ is the Eddington mass accretion rate.

To observe the maximum effect of the disk drag on the companion BH, we consider the orbital plane of the companion coincides with that of the disk, i.e. the companion will remain always embedded within the disk. To examine such orbital motion, we developed a numerical code for integrating the geodesic equations of the companion and calculate the loss rates from the instantaneous 4-momentum ($P^{\mu}$) of the companion during its orbital evolution including all such effects. The only required initial parameters that remain for this code are the initial semi-major axis ($X$) and eccentricity ($e$), Kerr parameter ($a$), initial E/IMR ratio, mass accretion rate $\dot{m}$ and the disk parameters.

The orbital evolution of the companion is solely governed by the loss rates and the instantaneous values of the orbital parameters. The loss rates depend on the accretion rate, therefore, throughout our study, we took fixed accretion rate $\dot{m}=1.0$. With this configuration, the evolution of the orbital parameters is studied and its variations are observed during the transition from the inspiral to the merger stage.

\subsection{Effects of hydrodynamic drag}

\begin{figure}
\centerline{\includegraphics[width=1.0\linewidth,clip]{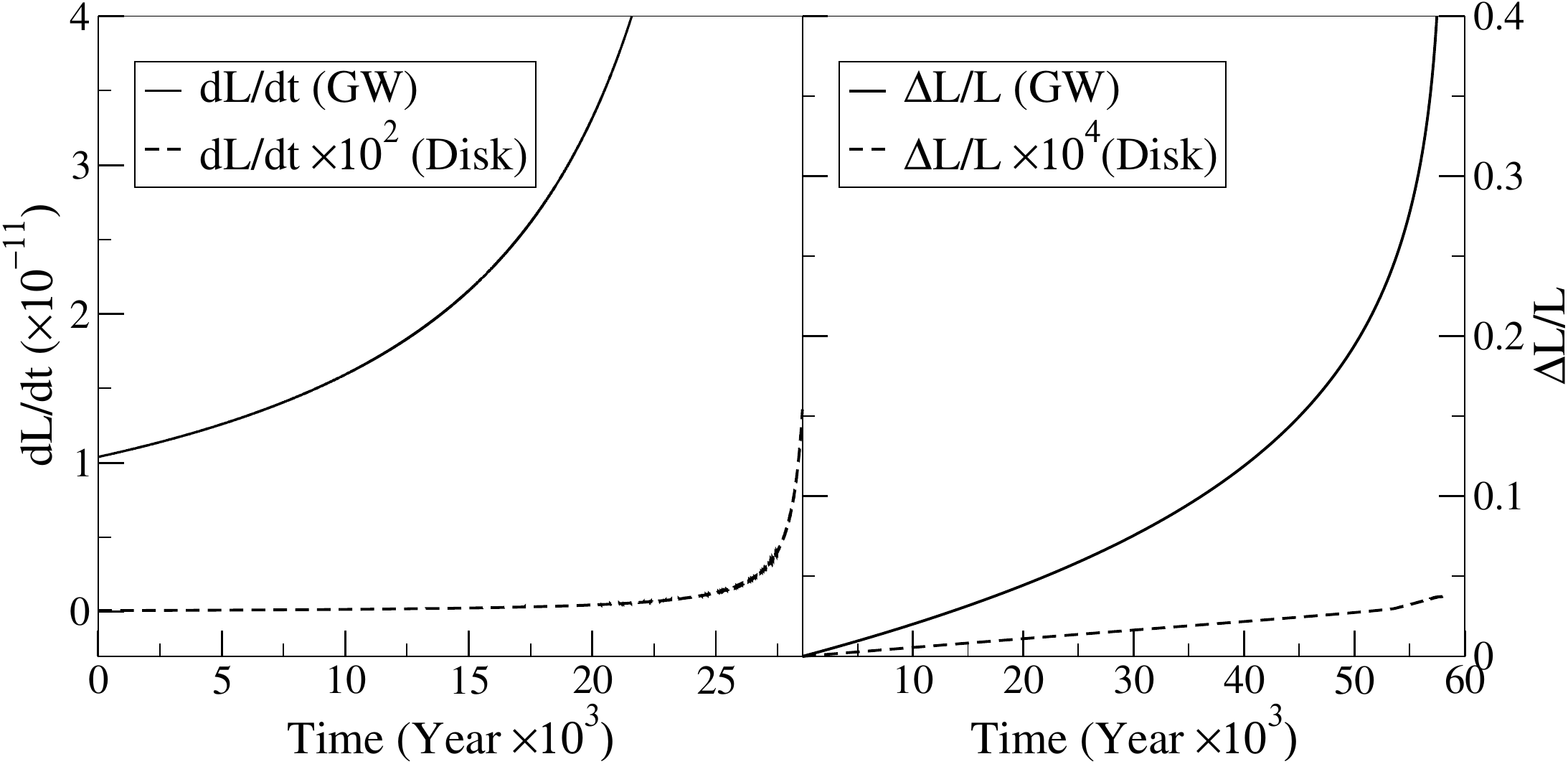}}
\caption{
\label{Lcom1} We plot the emission rate of angular momentum ($L$) of the companion black hole (left panel) due to the GW emission(\textit{solid line}) and due to the disk drag (\textit{dash line}) when companion moves through the accretion disk. In the right panel, we plot the instantaneous change of angular momentum of companion BH $\frac{\Delta L}{L}$ due to the GW emission(\textit{solid line}) and due to the disk drag (\textit{dash line}). In the left panel we choose EMRI with masses $M=10^5M_{\odot}$ $m=10M_{\odot}$ and for the right panel we choose $M=10^8M_{\odot}$ $m=10^6M_{\odot}$. Other parameters are $\dot{m}=1$, $a=0.50$, $e=0.30(left)$ $e=0.50(right)$ and  $X=500r_g$. From both panel we see that a small and non-negligible contribution is introduced by the disk over the GW loss rates. }
\end{figure}

In the presence of the disk, the emission rate due to the drag effect on the companion black hole has been compared with the emission rates of gravitational waves. In the left panel of the figure \ref{Lcom1}, we present the angular momentum emission rates due to GW (\textit{solid line}) and due to hydrodynamic drag (\textit{dashed line}). Due to these losses angular momentum of the companion changes. The change of the instantaneous angular momentum $\frac{\Delta L}{L}$ due to GW (\textit{solid line}) and due to drag(\textit{dashed line}) is plotted in the right panel of the figure. From the figure, we see that hydrodynamic drag is much smaller than that due to gravitational wave.

\begin{figure}
\centerline{\includegraphics[width=0.9\linewidth,clip]{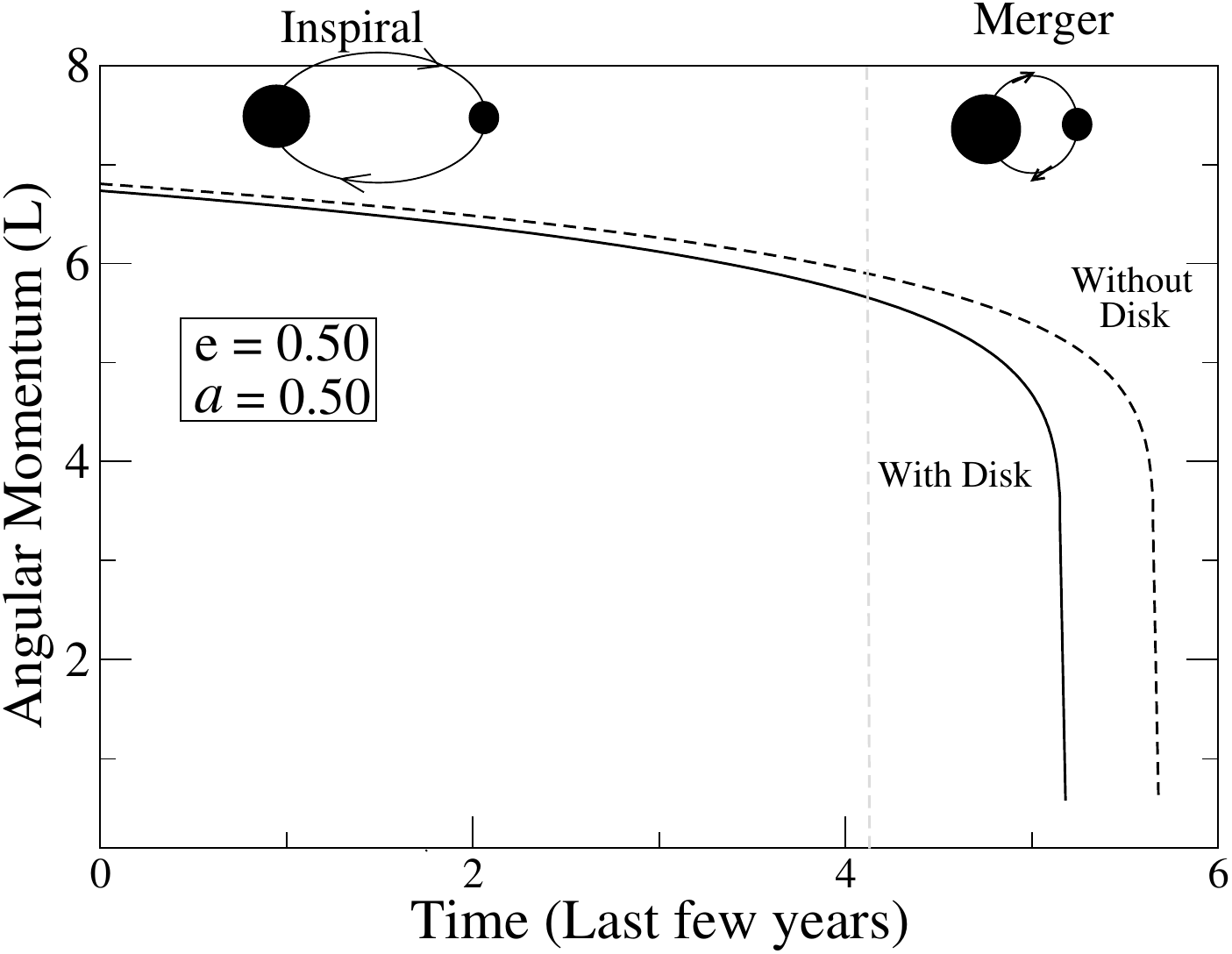}}
\caption{
\label{LtimeGWdisk} In the figure, we compare the angular momentum ($L$) loss of the companion with time ($T$) when the accretion disk is present (\textit{solid line}) and absent (\textit{dash line}). We plot the figure for the last few years before coalescence, where we can see the change of $L$ ( between solid and dashed line), increases from the inspiral to merger stage. We see in the presence of the disk coalescence takes place comparatively faster ($\sim$ 1 year) than in the scenario of disk absence. Initial parameters  are: $\dot{m}=1$, $e=0.50$, $a=0.50$, $X=500r_g$,$M=10^8M_{\odot}$, $m=10^6M_{\odot}$.}
\end{figure}

These changes are observed for various types of orbits. For instance, we consider elliptic orbits with different initial eccentricities and different semi-major axes.  In Figure \ref{LtimeGWdisk}, we plot the change in angular momentum ($L$) during the orbital evolution in the last 6 years before the coalescence. We present our result with initial eccentricity and Kerr parameters value $e=0.50$ and $a=0.50$  respectively for two cases : (i) GW+Disk (\textit{solid line}) and (ii) GW only (\textit{dashed line}). We see that in the presence of the disk, the angular momentum decreases comparatively faster than without the disk. Thus the coalescence time is shorter by almost one year. The coalescence time reduces further for high eccentric orbits.

\begin{figure}
\centerline{\includegraphics[width=1.0\linewidth,clip]{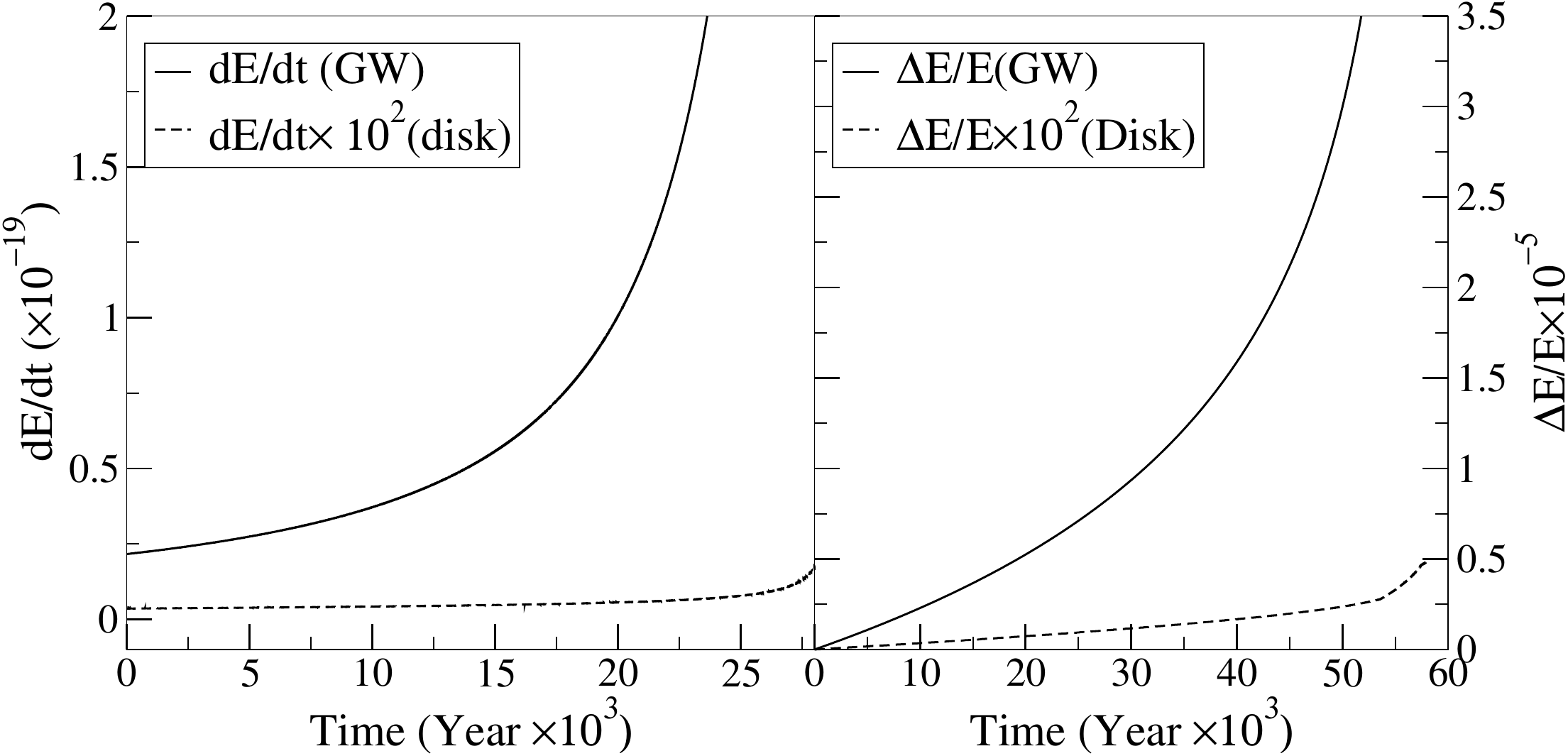}}
\caption{
\label{Ecom1} In the left panel we plot $\dot{E}$ vs time for the system with central BH mass $10^5M_{\odot}$ and $EMR=10^{-4}$ and in the right panel we plot $\frac{\Delta E}{E}$ for a system with companions mass $10^8M_{\odot}$ and $10^6M_{\odot}$. In both panels of the figure, we present a comparison of the energy loss of companion black holes between the loss rates due to the emission of GW(\textit{solid line}) and due to the disk drag(\textit{dash line}) while companion BH($m$) moves through the accretion disk. Here again, it clearly shows that the energy loss is small but non-negligible in the presence of the disk.  Other parameters are the same as figure \ref{Lcom1}.}
\end{figure}

A very similar change is noticed in the energy loss of the companion. In the left panel of the figure \ref{Ecom1}, we present the energy emission rates due to GW (\textit{solid line}) and due to hydrodynamic drag (\textit{dashed line}). Due to these loss rate the change of the energy of the companion $\frac{\Delta E}{E}$ due to GW (\textit{solid line}) and due to drag(\textit{dashed line}) is plotted in the right panel of the figure. Here again, we see that the energy loss rate due to hydrodynamic drag is much smaller than the gravitational wave. These changes are also observed for various types of orbits considered before.

\begin{figure}
\centerline{\includegraphics[width=0.9\linewidth,clip]{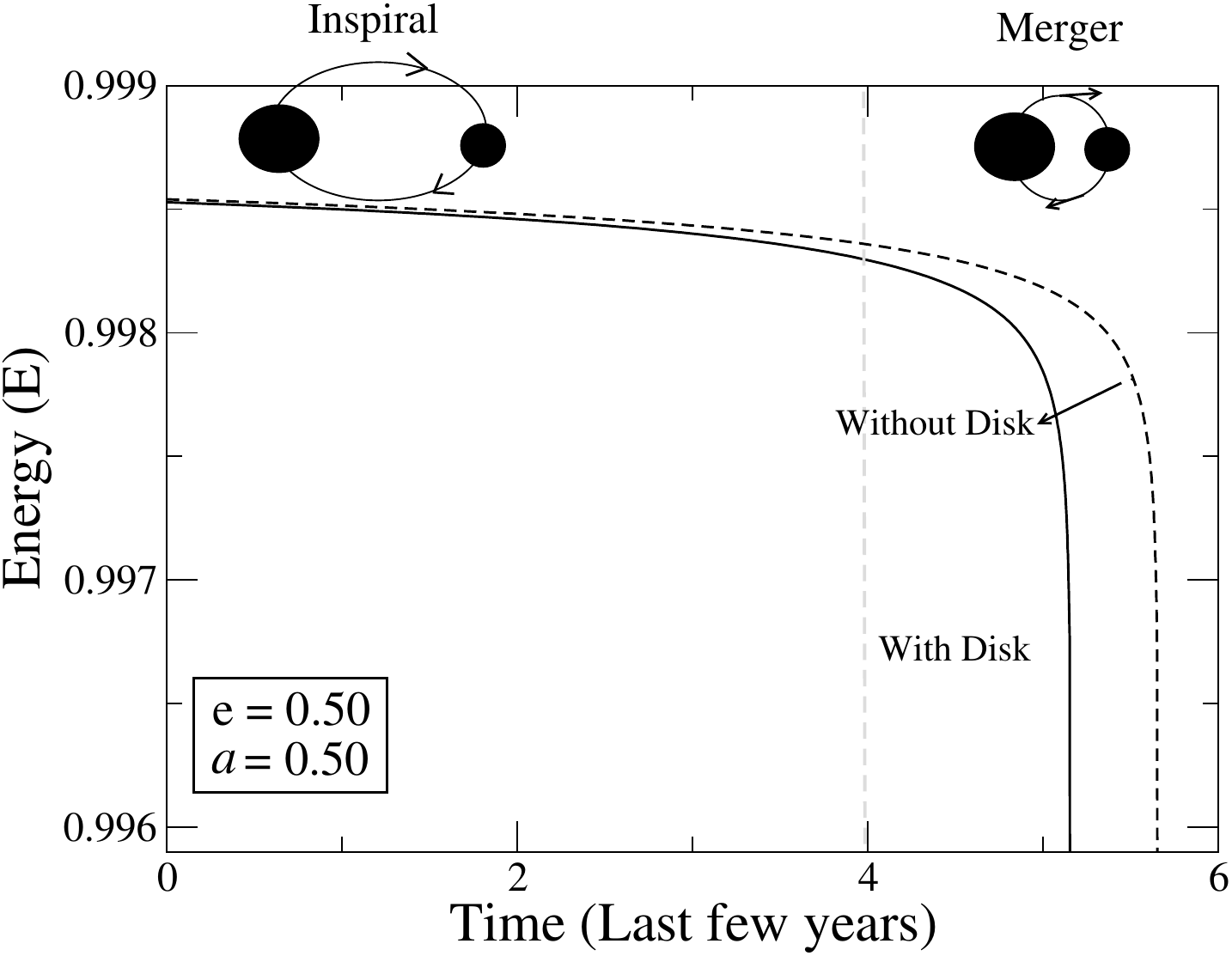}}
\caption{
\label{EtimeGWDisk} We compare the results of the decrease in the energy ($E$) of the companion black hole when the accretion disk is present (\textit{solid line}) and absent (\textit{dash line}).  We plot the figure for the last few years before coalescence, where we can see the change of $E$ ( between solid and dashed line), increases when EMRI moves from the inspiral to merger phase. Though the change is small, we see that in the presence of the disk, the companion loses energy faster($\sim$ 1 year) than in the absence of the accretion disk. Other parameters are the same as mentioned in figure \ref{LtimeGWdisk}. }
\end{figure}

In figure \ref{EtimeGWDisk}, we plot the variation of energy in the last 6 years before the coalescence. In the figure, we see that, in the presence of the disk, the energy loss rate is comparatively higher than in the absence of the disk indicating that a non-negligible hydrodynamic drag acted on the companion due to the disk. The reduction of coalescence time in present (\textit{solid line}) and absence (\textit{dashed line}) of the disk is consistent with that of figure \ref{LtimeGWdisk}. For high eccentric orbits, this coalescence time further reduces.

From the above numerical results, we now conclude that energy and angular momentum loss is increased in the presence of the disk thereby changing the orbital motion of the companion. In comparison to the GW emission, a small but non-negligible amount of angular momentum loss is contributed by the disk.

\subsection{Torque on the Companion}
We have calculated $\frac{dL}{dt}$ in the presence and absence of the accretion disk respectively. The difference between their values gives the torque due to the disk only (i.e., hydrodynamic drag ). We compare the ratio of the torque due to GW and Disk drag acted on the companion in the late inspiral stage (radius between $10r_g$ to $5r_g$). We find that for the system with masses $(10^6M_{\odot},10^3M_{\odot})$ the ratio of the torque (GW to Disk) is $\sim$ $10^{2}-10^{3}$ for our case and $10^{3}-10^{5}$ for other models (e.g. \citet{2019MNRAS.486.2754D}). However for the EMRIs $(10^5M_{\odot},10^2M_{\odot})$ and  $(10^5M_{\odot},10M_{\odot})$ the ratio becomes $\sim$ $10^{3}-10^{4}$ and $\sim$ $10^7$ respectively. From our study we found that, for our chosen system parameters, our drag force is higher than what estimated in other studies. Also for other system parameters it could be less. The net accumulated effect of the drag will be manifested on the dephasing of GW signal. We also measured and compared with other studies in the upcoming section \S\ref{dephs1} and \S\ref{snr11}.

\subsection{Evolution of the orbital parameters in the presence and absence of the disk}

\begin{figure}
\centerline{\includegraphics[width=0.9\linewidth,clip]{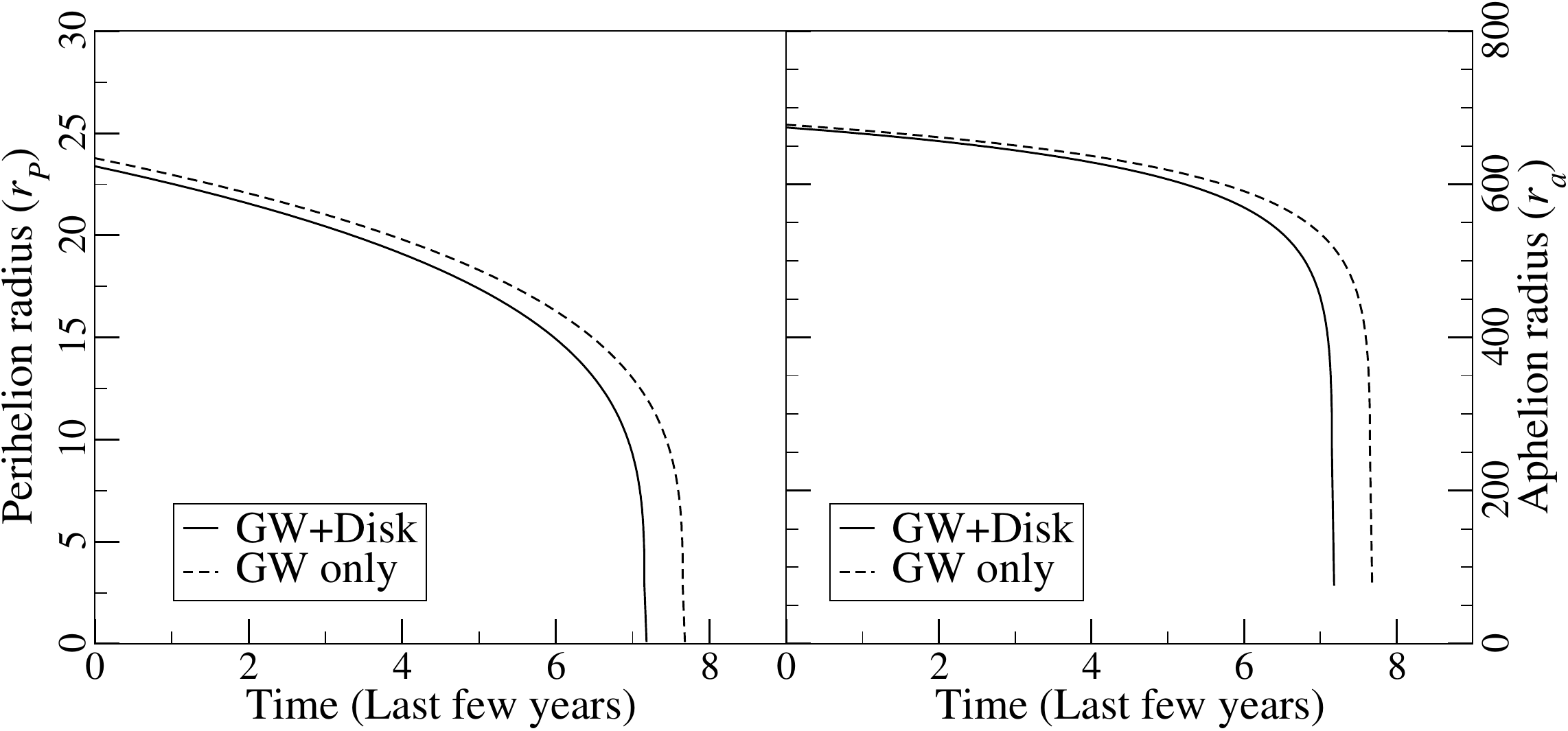}}
\caption{
\label{rTGWD} In this figure we compare the decrease of perihelion ($r_P$) and aphelion ($r_a$) radius with time ($T$) for the two cases when accretion disk is present (\textit{solid line}) and when the disk is absent (\textit{dash line}). We observe the changes for the last few years before the coalescence. We see in the presence of the disk both $r_P$ and $r_a$ of the orbit reduce faster($\sim$ 1 year) than in the absence of an accretion disk. Initial parameters considered for this plot are as follows: $\dot{m}=1$, $e=0.50$, $a=0.50$, $X=500r_g$, $M=10^8M_{\odot}$, $m=10^6M_{\odot}$.}
\end{figure}

The loss of $E$ and $L$  causes the orbital radius gradually decreases and results in an inspiral motion. Therefore the orbits change with time and the instantaneous trajectory depends on the instantaneous values of $E$ and $L$ . In the case of an elliptical orbit or inspiral, a decrease in the perihelion and aphelion radius is thus calculated. In Figure \ref{rTGWD}, we present the decrease in both perihelion ($r_P$) and aphelion ($r_a$) radius with time ($T$) in the presence and the absence of the disk. Attention is made particularly in the last few years before coalescence when the companion moves to the inspiral to merger stage. The initial values of the orbital radii and other orbital specifications are given in the figure. We see both the aphelion and perihelion distances reduce similarly with time. However the differences between the gradual change of radial parameters with and without a disk are noticeable and in the presence of the disk, the radii decrease at a faster rate. 

\begin{figure}
\centerline{\includegraphics[width=0.9\linewidth,clip]{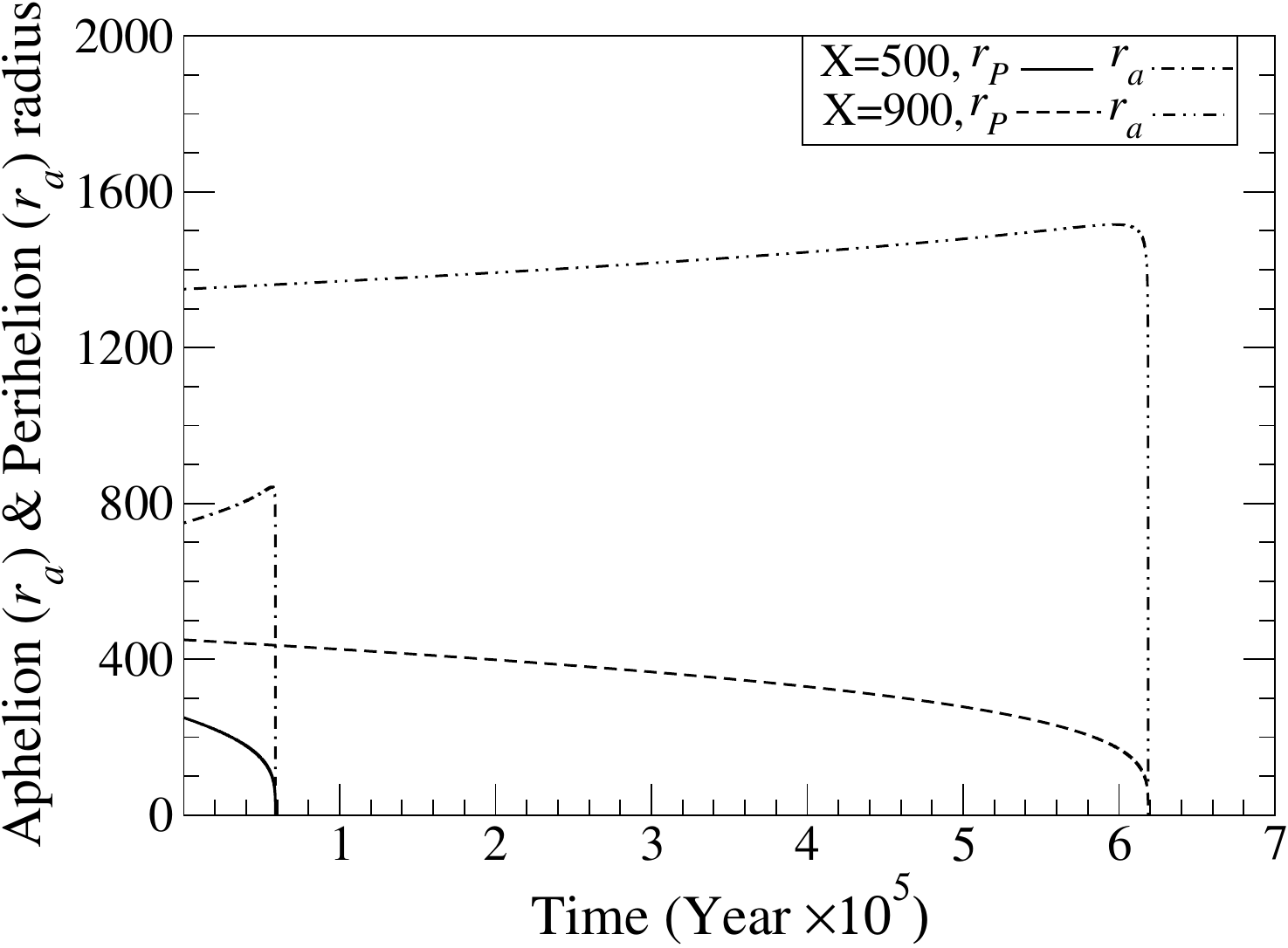}}
\caption{
\label{rarpT59} A comparison of the perihelion ($r_P$) and aphelion ($r_a$) radius with time ($T$) for two orbits with semi-major axis $X=500r_g$ and $X=900r_g$ is shown in this figure. The $r_P$ and $r_a$ for $X=500r_g$ are shown by \textit{solid line} and \textit{dot-dash line}. For $X=900r_g$, the $r_P$ and $r_a$ are shown by \textit{dash line} and \textit{dot-dot-dash line}. As expected, from the figure we see that the $X=900r_g$ orbit takes a larger time for circularization ($\sim 0.5$ million years). Initial parameters used here are : $\dot{m}=1$, $e=0.50$, $a=0.50$, $M=10^8M_{\odot}$, $m=10^6M_{\odot}$.}
\end{figure}

The total infall time before the coalescence is shown in Figure \ref{rarpT59}. Here we consider two different orbits with initial values $X=500r_g$ ($r_P$ (\textit{solid line}) and $r_a$ (\textit{dot-dash line})) and $X=900r_g$ ($r_P$ (\textit{dash line}) and $r_a$ (\textit{dot-dot-dash line})) and observe the decrease in perihelion ($r_P$) and aphelion ($r_a$) radius. The $r_p$ decreases at a faster rate however $r_a$ slowly increases in the inspiral stage. The orbit evolves in such a way that the eccentricity of the orbit decreases very slowly with time to make it circular. In the final merging stage before the coalescence, both radii sharply fall. 

\begin{figure}
\centerline{\includegraphics[width=1.0\linewidth,clip]{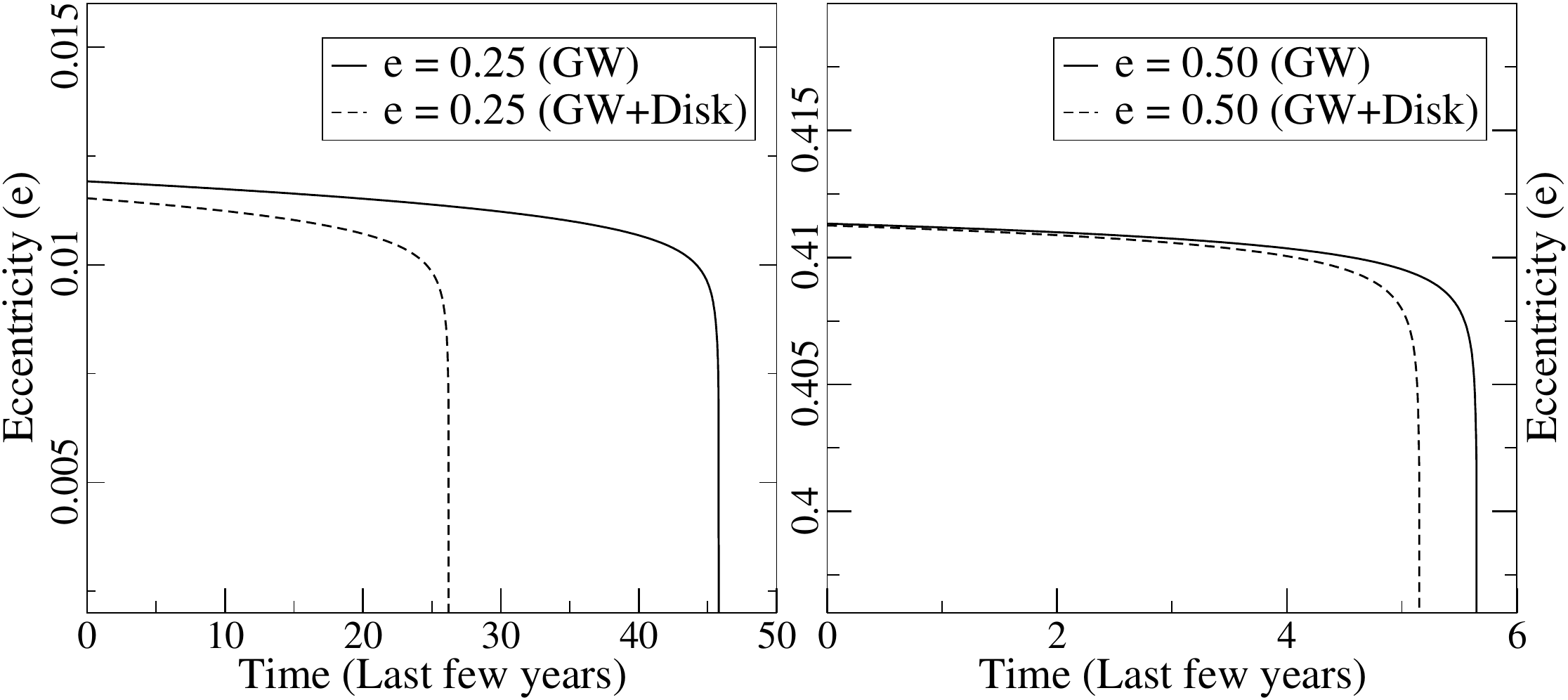}}
\caption{
\label{ecnTGW} This figure shows the decay of eccentricity ($e$) of the orbit during the last few years of observation. We compare two eccentric orbits with initial eccentricities $e=0.50$ (high) and $e=0.25$(moderate). In both cases, when the accretion disk is absent, ( is shown by \textit{solid line}) and when the disk is present ( shown by \textit{dash line}). We see that in the presence of the disk, the companion's orbit loses eccentricity comparatively faster for $e=0.25$ than for $e=0.50$. This is due to the companion traveling longer inspiral for $e=0.25$. The initial parameters used for these plots are $\dot{m}=1$, $a=0.50$, $X=500r_g$, $M=10^8M_{\odot}$, $m=10^6M_{\odot}$.}
\end{figure}

Eccentricity can be expressed interms of $r_a$ and $r_p$ as $e(T)=\frac{r_a-r_p}{r_a+r_p}$. During GW emission, a circular orbit (with eccentricity $e=0$) of the companion remains circular and an elliptic orbit always tends to be circular by reducing its eccentricity. Our results are consistent with it. The decrement rate ($\dot{e}=\frac{de}{dT}$) is higher for high elliptic orbits. In our numerical studies, we have considered several elliptical orbits, and two such examples with initial eccentricities $e=0.50$ and $e=0.25$ are presented in Figure \ref{ecnTGW}. In both cases, when the accretion disk is absent is shown by (\textit{solid line}), and when the disk is present is shown by (\textit{dash line}). The Kerr parameter for the two cases is $a=0.50$. Here again, the evolution is shown in the last few years before the coalescence. The presence of the accretion disk reduces the eccentricity further hence making it circular faster. However the low eccentric orbit spares a longer time, therefore, the effect of the hydrodynamic drag is large compared to the high eccentric orbits. The separation of the solid and dotted lines is large for $e=0.25$ and less for the $e=0.50$. From the figure we see that in the case of eccentric orbit $e=0.25$ coalescence time reduces significantly around 20 years in the presence of the disk while for the orbit with $e=0.50$ very less reduction in coalescence time ($0.5$ years) occurs.

\begin{figure}
\centerline{\includegraphics[width=0.9\linewidth,clip]{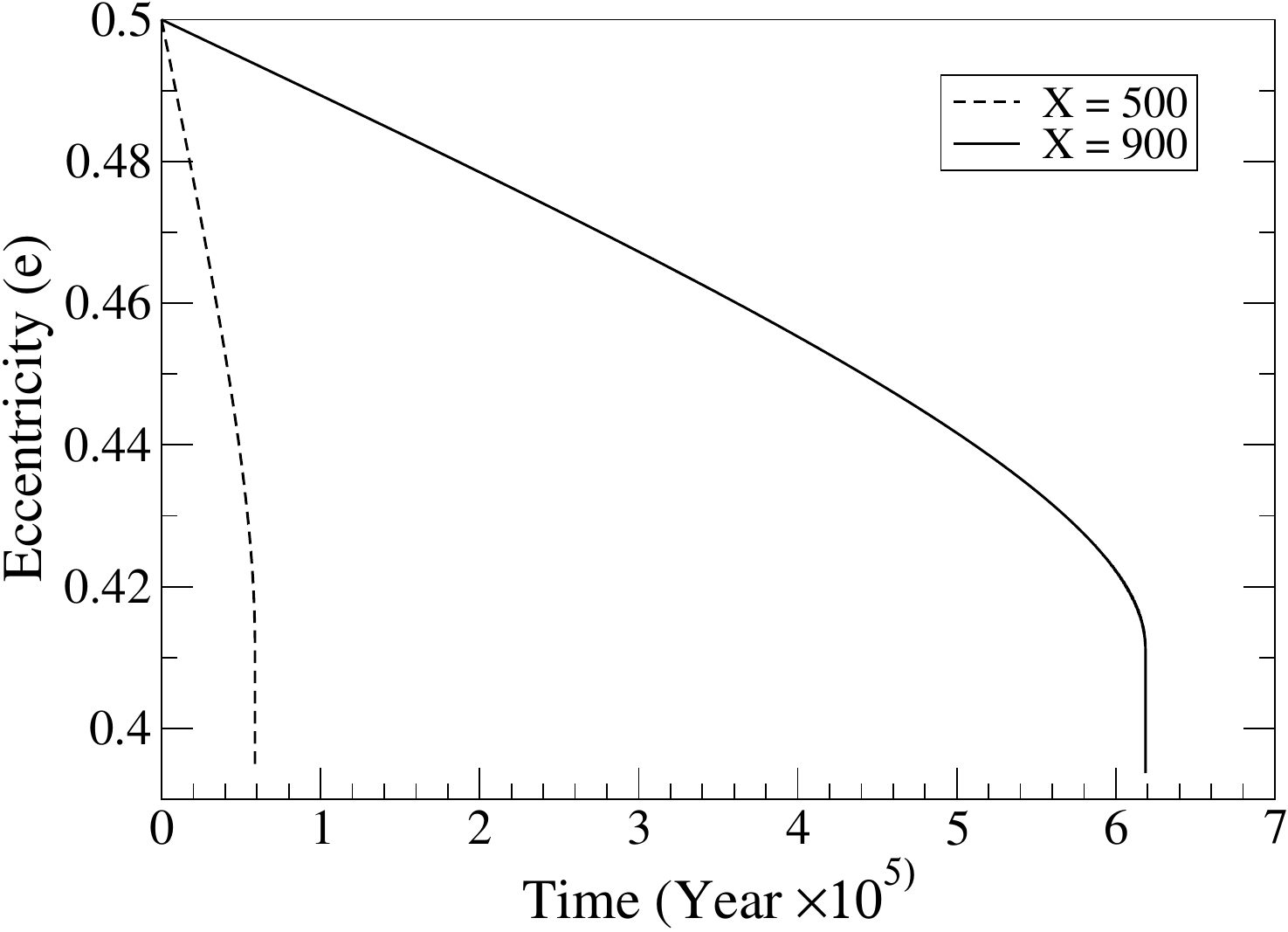}}
\caption{
\label{ecnT59} In this figure we plot the variation of eccentricity($e$) for two orbits with semi-major axis $X=500r_g$ (\textit{dash line}) and $X=900r_g$(\textit{solid line}). As expected the companion BH took a smaller time for the orbit with a low semi-major axis.   Parameters used are $\dot{m}=1$, $e=0.50$, $a=0.50$, $M=10^8M_{\odot}$, $m=10^6M_{\odot}$.}
\end{figure}

The eccentric orbit with initial eccentricity $e=0.50$ further investigated for different initial parameters having semi-major axis $X=500r_g$(\textit{dash line}) and $X=900r_g$(\textit{solid line}) in the presence of the disk. Here in Figure \ref{ecnT59}, we see that as usual the larger orbit decay slower and takes a longer time for migration. Therefore the eccentricity evolves slower for $X=900r_g$ and faster for $X=500r_g$. However, the amount of change of eccentricity is the same for both cases even though the inspiralling stages are different.

\begin{figure}
\centerline{\includegraphics[width=0.9\linewidth,clip]{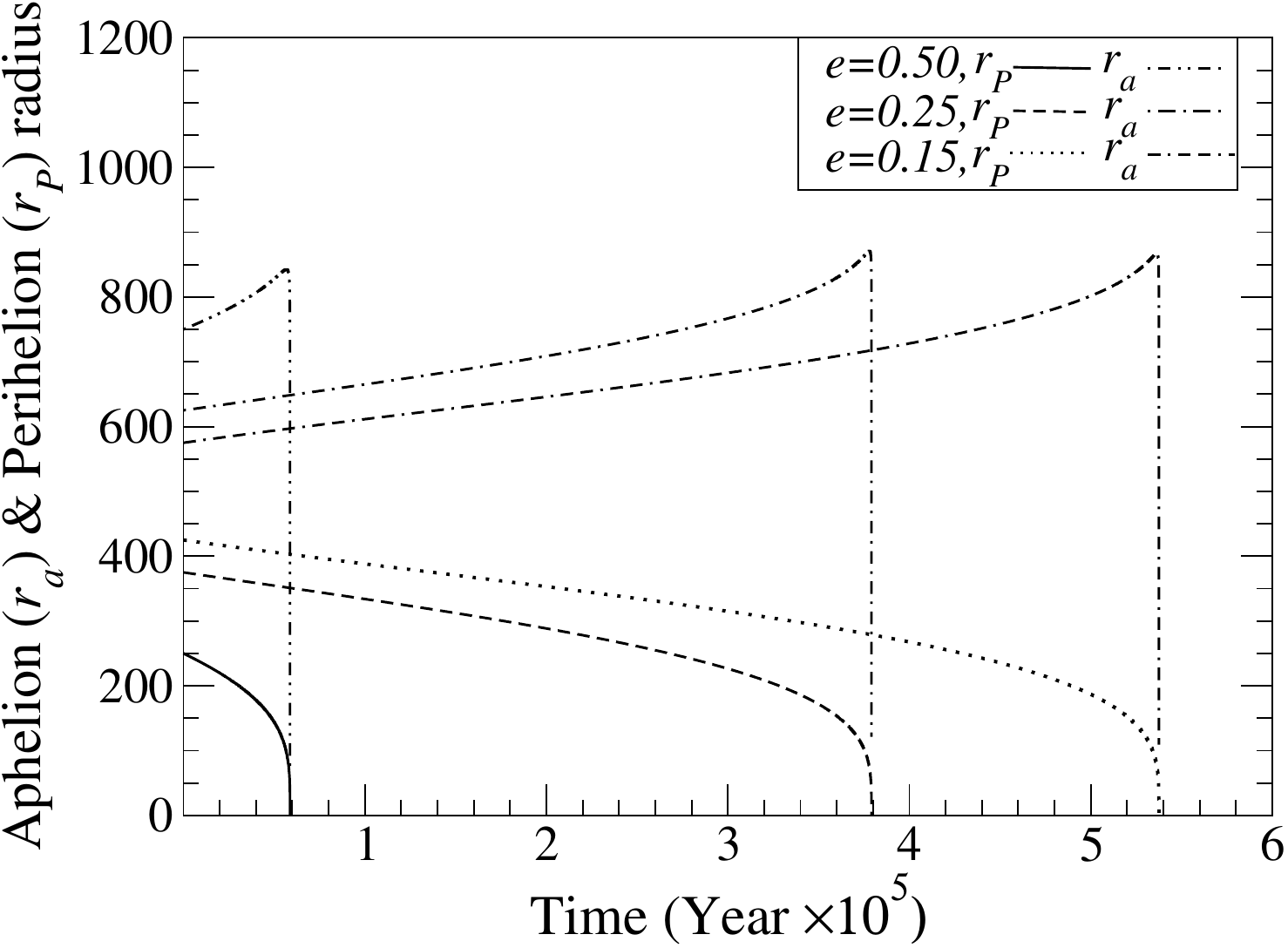}}
\caption{
\label{rarpTecn} In this figure, we plot the variation of the perihelion ($r_P$) and aphelion ($r_a$) radius of the companion black hole's orbit with time ($T$) for three cases with eccentricities $e=0.15$, $e=0.25$ and $e=0.50$. From the figure, we see that $r_a$ initially increases and after a while decreases more rapidly. The circularization takes place faster for the comparatively high eccentric orbit. The initial parameters are $\dot{m}=1$, $a=0.50$, $X=500r_g$, $M=10^8M_{\odot}$, $m=10^6M_{\odot}$. }
\end{figure}

The total infall time for different eccentric orbits has been investigated in Figure  \ref{rarpTecn}. Here we plot the change of the perihelion ($r_P$) and aphelion ($r_a$) radius with time for different eccentric orbits having the same initial value of the semi-major axis $X=500r_g$. We compare the result for three orbits with low, moderate, and high eccentricities e.g. $e=0.15$, $e=0.25$, and $e=0.50$ having perihelion radii $r_P=425r_g$, $375r_g$, $250r_g$ respectively. Among the three the separation is more for $e=0.15$, $r_P=425r_g$, and therefore the companion travel a longer distance through the disk and hence took a longer time before coalescence. The coalescence time reduces to 0.15 million years for the orbit with $e=0.25$ and $r_P=375r_g$. The coalescence time significantly reduces around 0.5 million years for high eccentric orbit $e=0.50$ and $r_P=250r_g$. Hence the infall time is reduced significantly for high eccentric orbit.  Since low eccentric orbit took a longer time to become circular before coalescence, therefore the rate of changes of $r_a$ and $r_P$ is smaller in this case. The total orbital impact of hydrodynamic drag is much more (as it takes longer time) for low eccentric orbit however the coalescence takes place faster for high eccentric orbit. The decay of the orbits can be understood from the variation of the $r_P$ values for three cases as shown in Figure \ref{rarpTecn}.

\begin{figure}
\centerline{\includegraphics[width=0.9\linewidth,clip]{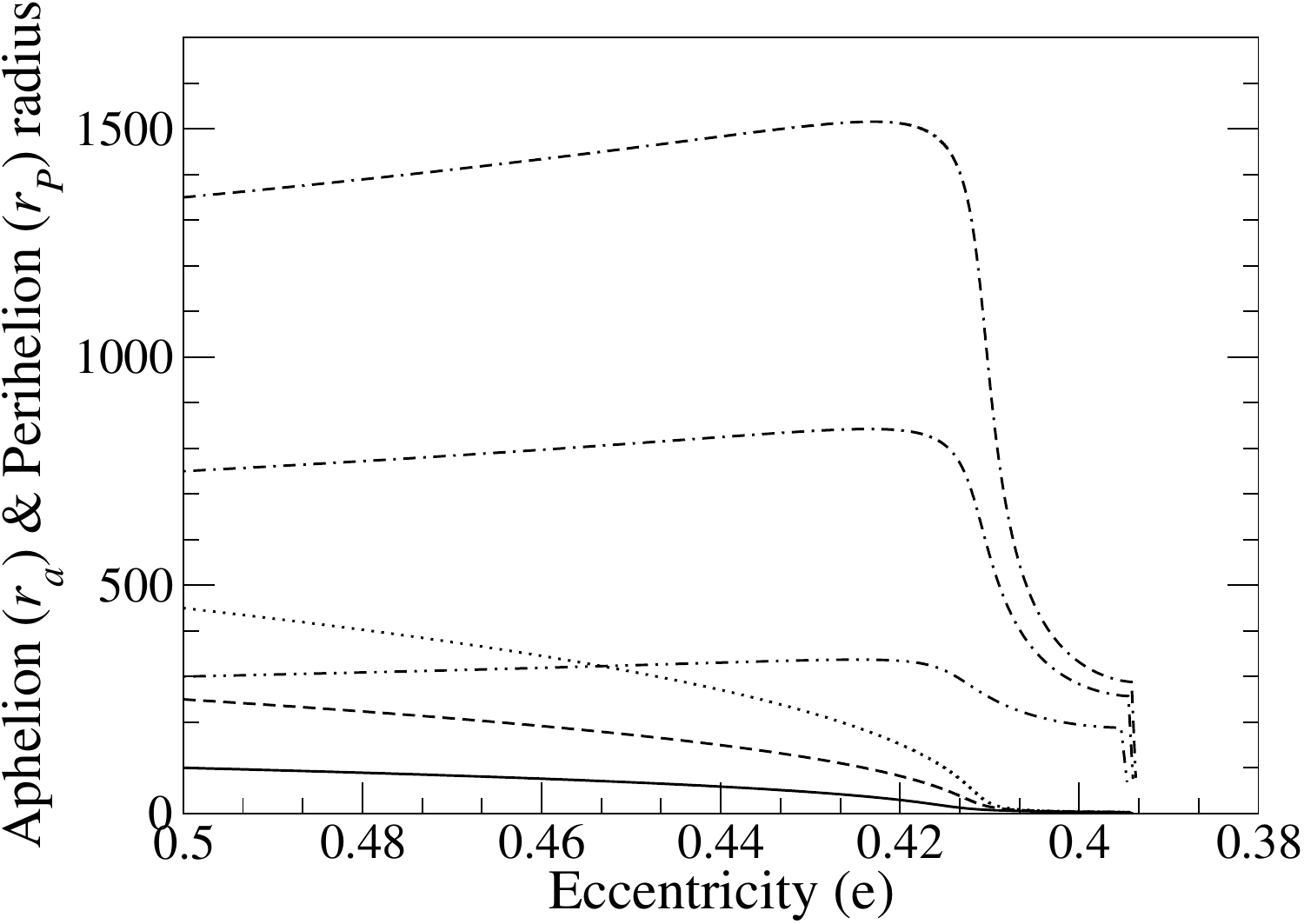}}
\caption{
\label{ecnrarp259} A comparison of the perihelion ($r_P$) and aphelion ($r_a$) radius with the eccentricity ($e$) of the orbit is presented in the figure. Initial eccentricity $e=0.50$ is considered for all of them. The different semi-major axis $X=200r_g$ ($r_P \rightarrow $ \textit{solid line},$r_a \rightarrow $ \textit{dot dot dash line}), $X=500r_g$ ($r_P \rightarrow $ \textit{dash line},$r_a \rightarrow $ \textit{dot dash line}) and $X=900_g$ ($r_P \rightarrow $ \textit{dot line},$r_a \rightarrow $ \textit{dot dash dash line}) respectively are shown. In all three cases, eccentricity is reduced to almost the same value.}
\end{figure}

The variation of $e$ for high eccentric orbit is investigated for different initial configurations. We choose three different values of the initial inspiral radius(low, moderate, and high) of the binary mergers having semi-major axes $X=200r_g$, $X=500r_g$, $X=900r_g$ and plot the variation of perihelion ($r_P$) and aphelion ($r_a$) radius with the eccentricity ($e$) in the Figure \ref{ecnrarp259}. In three such cases, the initial eccentricity is $e=0.50$. In Figure \ref{ecnT59}, we already noticed that coalescence time is larger for $X=900r_g$ and smaller for $X=500r_g$. The coalescence time is even smaller when we choose $X=200r_g$. Even though three different stages of inspiral with the same eccentricity initially started, we observe an almost equal change in eccentricity for all three cases. The reason could be the disk drag is more effective near the central accreting object because of the high radial velocity of the disk in which the companion is moving around.

\subsection{Dephasing of the companion}
\label{dephs1}
As can be seen from the previous plots the loss of $E$ (figure \ref{EtimeGWDisk}) and $L$ (figure \ref{LtimeGWdisk}) due to disk is small compared to GW radiation however at the end of a long-term inspiral, the accumulated effect of the drag is finite. Further, we see in the figure \ref{EtimeGWDisk} and \ref{LtimeGWdisk} that, there is a total time lag in the infall time.  Although this time lag is small compared to the total journey time starting from the early inspiral stage, however here again, during the late inspiral stage, the accumulated effect of this time lag creates a definite phase lag which may be detectable in LISA \citep{2019MNRAS.486.2754D,2019MNRAS.489.4860D,2021MNRAS.501.3540D,2022arXiv220710086S}. 
We, therefore, calculate the phase acquired by the companion with and without disk separately and measure the phase difference ($\delta \phi$).

\begin{figure}
\centerline{\includegraphics[width=0.9\linewidth,clip]{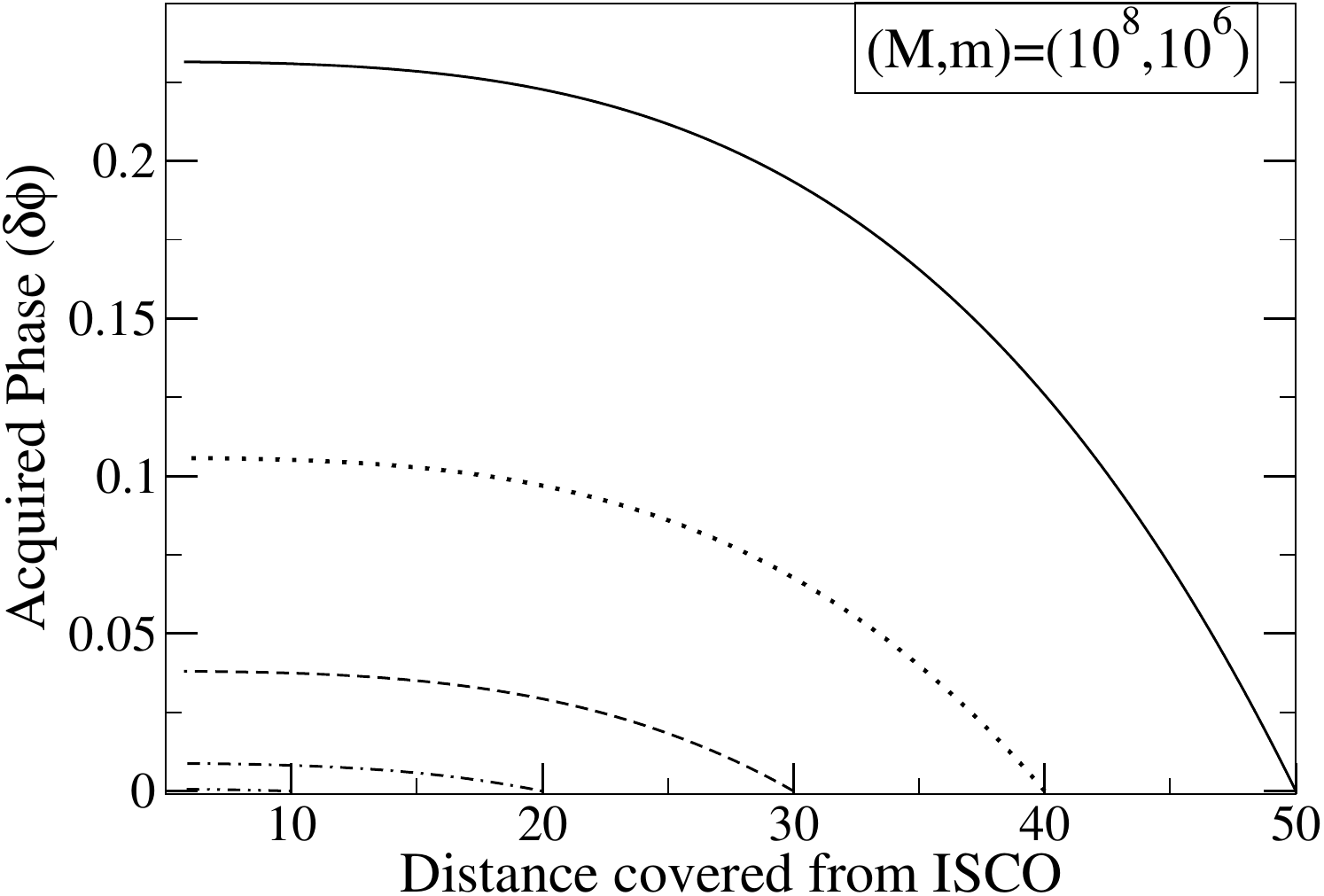}}
\caption{
\label{phase} In this figure, we plot the total phase difference acquired by the companion BH while it moves through the disk. The \textit{solid line} represents the amount of $\delta \phi$ gained by the companion in the last 5 years before merging. Respective phase accumulation is plotted for smaller separation covered from ISCO e.g. for $40r_g$(\textit{dotted line}), $30r_g$(\textit{dashed line}), $20_g$(\textit{dot-dashed line}), $10r_g$(\textit{dot-dot-dashed line}). We see that the phase difference $\delta \phi$ increases monotonically during orbital decay. The curve shows a plateau
at the late stage of inspiral and common even when the separation is smaller. However, the magnitude of the plateau decreases gradually. The detectability of these phase shifts obtained during different phases of observation would be known by estimating the signal-to-noise ratio(SNR) of the sources and thus may be detectable by the present-day GW detector LISA.}
\end{figure}

We plot this phase shift ($\delta \phi$) with radius for different binary separations in Figure \ref{phase}. The figure gives an estimation of the phase difference acquired by the companion during the last 5 years before the coalescence when the separation between them was close enough and typically around 50$r_g$, where $r_g$ is the Schwarzchild radius. We see that the phase difference increases monotonically during the orbital decay from $50r_g$ onwards. The curve shows a plateau at the late stage of the inspiral. This plateau is common when the separation is smaller however the amount of phase shift reduces gradually. Hence depending on the length of the observation period the amount of phase shift would be greater or less. Whether this phase shift is detectable or not can be known by estimating the signal-to-noise ratio (SNR).

If the SNR is above the detectable range, this result could be interesting because the existence of the disk would be then identifiable through the accumulated phase shift during GW observation.

\begin{figure}
\centerline{\includegraphics[height=5cm,width=9cm]{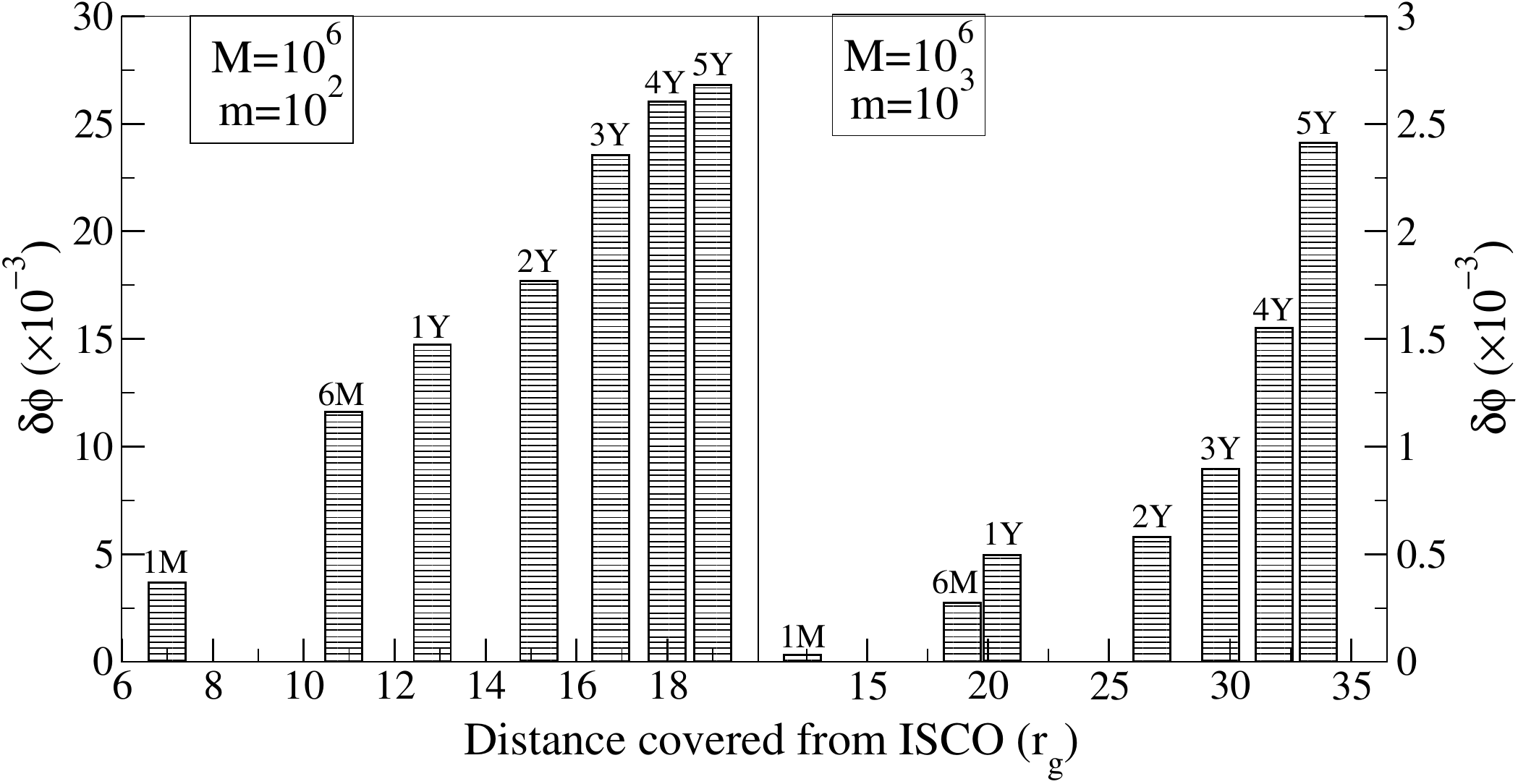}}
\caption{
\label{phsecom} {In this figure we plot the different histograms of the acquired phase shift $\delta \phi$ for different periods of observation. The observed period is calculated for different separations estimated from ISCO. We have plotted two different cases of E/IMRI in panels (a) \& (b). From both panels, we see that as the observation period or the orbital radius decreases the accumulated phase shift due to disk drag decreases. Here again, the detectability of these phase shifts during one particular observation period would be known from SNR values.}}
\end{figure}

From the result we see that the dephasing of the companion black hole depends on the period of observation. To find its detectability during different phases of observation we consider two different systems with mass ratio $EMR=10^{-4}$ and $10^{-3}$ having central BH mass $10^6M_{\odot}$ at a distance of 1 GPc from the detector. We calculate the dephasing of the companion in figure \ref{phsecom} when the disk is present and absent. We consider different observation times starting from the innermost stable orbit to calculate the accumulated phase shift for both cases. From the different histogram plots here again we see that the accumulated phase shift is different in every case as it depends on the separation between the companions of the system and also on the mass ratio. We found that the accumulated phase shift is detectable up to 6 months for $EMR=10^{-4}$ (left panel) however the detectability is less and a large observation period of about 5 years or more is required for the $EMR=10^{-3}$ (right panel).

Since we have used the transonic accretion flow model in our analysis (which has not been studied before for EMR cases), in table \ref{pcom}, we also compare the phase shift obtained by other disk models with our transonic disk model.

\begin{table}[H]
\centering
\begin{tabular}{|c|c|c|}
\hline
\multicolumn{3}{|c|}{Comparison of phase shift in last 1 year} \\
\hline
Mass Ratio  & Other Model & Transonic Disk\\
 \hline
 $(10^5,10)$ &$1.0 \times10^{-3}$ &  $1.91\times10^{-2}$\\
 \hline
 $(10^6,10^2)$ &$1.0\times10^{-3}$ &  $1.47\times10^{-2}$\\
    \hline
    \end{tabular}
    \caption{Comparison of GW dephasing produced by transonic accretion disk with the other disk model \citep{2011PhRvD..84b4032K,2019MNRAS.486.2754D}}
    \label{pcom}
\end{table}

We see that the dephasing is considerably higher than that calculated in recent simulation studies \citet{2019MNRAS.486.2754D,2011PhRvD..84b4032K} (at least for these cases mentioned in the table). Hence the dephasing of the companion solely depends on the choice of the accretion disk model. In our case that may be due to the difference in the non-Keplerian angular momentum and a high radial inflow velocity of transonic flows.

\subsection{Signal to Noise Ratio}
\label{snr11}
The acquired dephasing of the companion during different periods of observation would be detectable or not can be understood by calculating the SNR \citep{2011PhRvD..84b4032K,2015CQGra..32a5014M}. Since SNR depends on the mass ratio, source distance, and period of observation, thus we vary all these parameters and calculate the SNR for different cases, and presented them in a histogram diagram. We consider the detectability threshold of SNR $\ge 8$. When it is less than 8, we use a black bar to represent the non-detectability and the gray cylindrical bar for detectability (SNR$\ge 8$).

\begin{figure}
\centerline{\includegraphics[height=5.5cm,width=9cm]{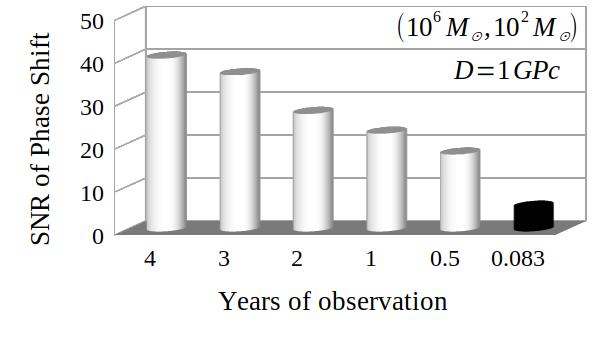}}
\caption{
\label{snr62y} {The histogram of SNR of the accumulated phase shift is plotted for the EMRI ($10^6M_{\odot},10^2M_{\odot}$) at a distance of 1 GPc away during the different observation period. From the figure, we see that SNR monotonically decreases as the observing period of LISA decreases. This is due to the fact that the accumulated phase shift is more for long-term observation (or for large separation). The gray cylindrical bars show a detectable SNR and the black bar represents non-detectability (SNR $<8$). The SNR is calculated upto the ISCO at $r_{ms}=4.23r_g$.}}
\end{figure}

In figure \ref{snr62y} we see that SNR of phase shift is calculated for different observational periods before merging for an EMRI of $M=10^6M_{\odot}$ and $m=10^2M_{\odot}$ located at a distance of 1GPc. From the figure, we see that the SNR of phase shift monotonically decreases with the period of observation and becomes non-detectable from the last month ($1month=0.083 year$) before merging. Prior to that, SNR is $\ge 8$ and the system is detectable.

\begin{figure}
\centerline{\includegraphics[width=1.0\linewidth,clip]{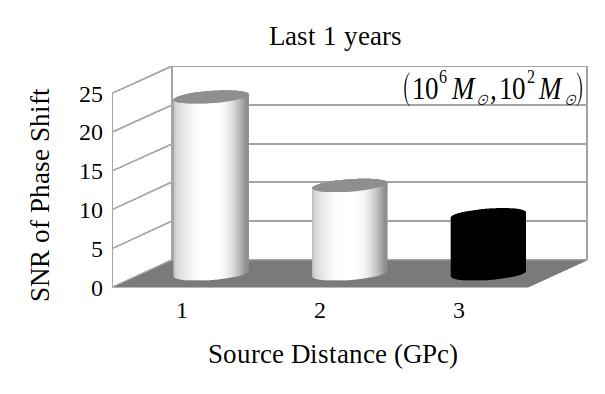}}
\caption{
\label{snr62D} {The SNR of the accumulated phase shift is plotted with the source distance D (GPc) for the last 1 year of observation. From the figure we see that as the EMRI to detector distance increases, the SNR inversely decreases with distance D. We also see that detectability decreases for far distance EMRI represented by the black cylindrical bar.}}
\end{figure}

The same EMRI varies with the source distance and we calculated the SNR of the phase shift in figure \ref{snr62D}. We see that when the source distance is 1GPc it is well within the LISA observational window however when we increase the source distance to 2GPc it is just detectable to LISA and remains undetectable when the distance is 3GPc or more.

\begin{figure}
\centerline{\includegraphics[width=1.0\linewidth,clip]{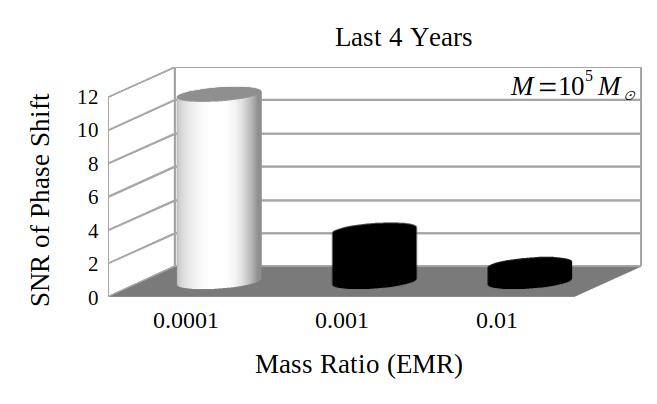}}
\caption{
\label{snr51} { The histogram of SNR of accumulated phase shift is plotted for different mass ratios $\frac{m}{M}$ of E/IMRI during the last 4 years of observation. The central SMBH mass $=10^{5}M_{\odot}$. From the figure, we see that the EMRIs(represented by gray bars) are more prominent and ideal sources for LISA detection than the IMRIs (represented by black bars). }}
\end{figure}

In the figure \ref{snr51}, we finally varied the mass ratio $EMR$ of the sources keeping the observation period and the source distance fixed. In the figure, we see that when the $EMR=10^{-4}$ the SNR of the phase shift is just detectable in the LISA band however when we decrease the mass ratio to $EMR=10^{-3}$ and $EMR=10^{-2}$ the system is not detectable. Hence we see that for 4 years of observational windows of LISA, the EMRIs are best detectable compared to the IMRIs.

\begin{figure}
\centerline{\includegraphics[width=1.0\linewidth,clip]{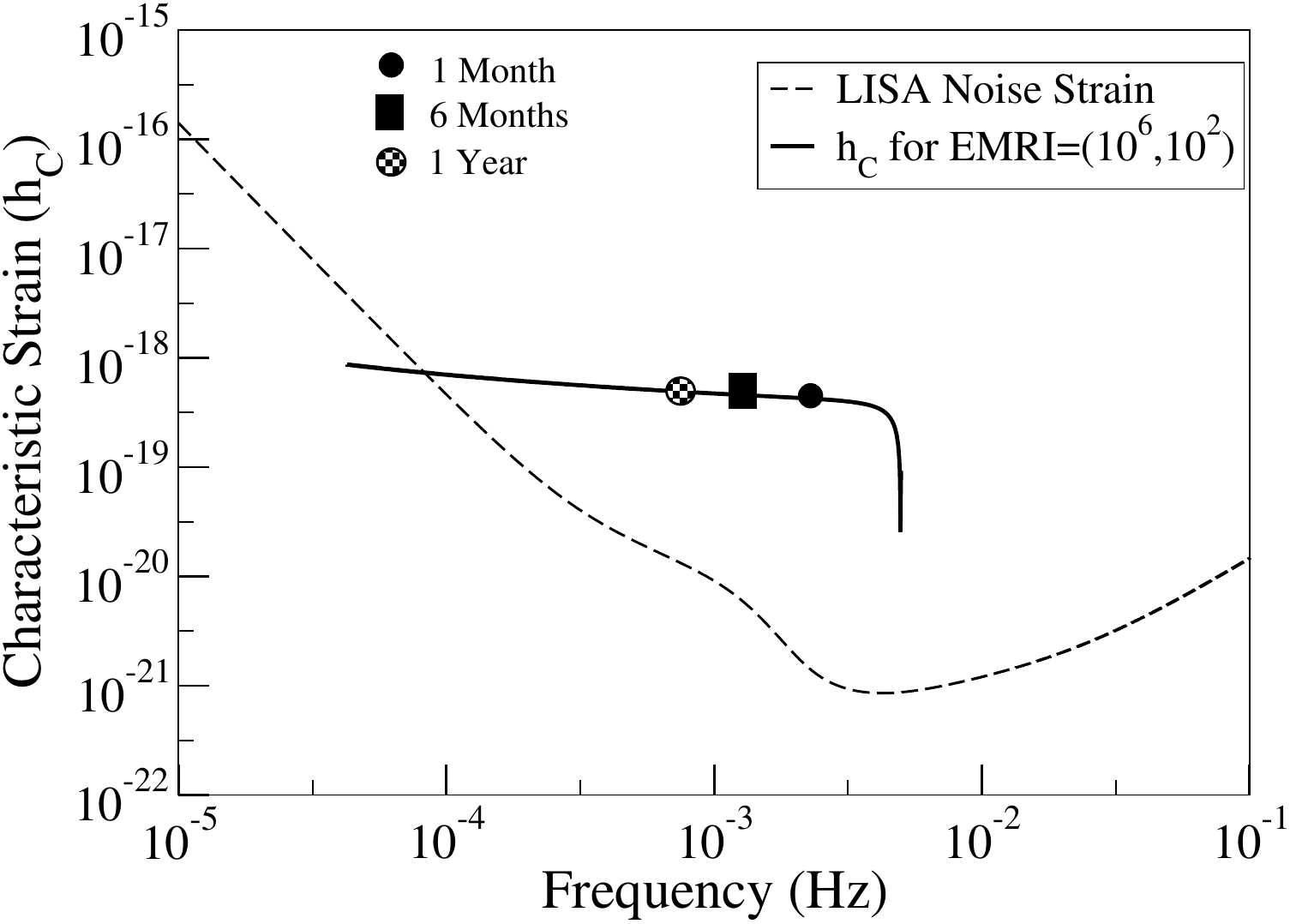}}
\caption{
\label{charS} {The \textit{dashed line} shows the LISA noise sensitivity curve with respect to the observed frequency of GW. Characteristic strain $h_c$ of EMRI ($M=10^{6}M_{\odot}$,$m=10^{2}M_{\odot}$) at a distance of 1 Gpc is plotted (\textit{solid line}) in the last five years of observation before merging. The strain is calculated up to marginally stable orbit ($r_{ms}=4.23r_g$) for $a=0.50$.  With time orbital radius shrinks and the frequency of the emitted wave enhances as indicated by the \textit{solid line} of the EMRI. We marked here when the companion BH is at 1 year, 6 months, and 1 month away before coalescence (denoted by different symbols as shown in the figure). From the figure, we see that the strain is so bright within the LISA window for a long time. This ensures a greater chance of observing the effect of disk drag by our transonic disk model. Thus the existence of the accretion disk could be identifiable in E/IMRIs and one may constrain the disk parameters from the observed GW data. }}
\end{figure}

In figure \ref{charS}, we present an interesting case where the EMRI could be detectable and well above the LISA noise strain. To represent that here we plot the characteristic strain $h_c$ for the EMRI (with central BH mass $10^{6}M_{\odot}$ and companion mass $10^{2}M_{\odot}$ at a distance of 1 GPc) during the last 5 years of observation before reaching the last stable orbit ($r_{ms}$) and the LISA noise strain (\textit{dashed line}) simultaneously. Using different symbols we locate the positions when the companions are 1 month, 6 months, and 1 year away from $r_{ms}$. With time frequency increases and so enhances the detection probability of the EMRI. Hence it is clear that the motion of the companion in the presence of a transonic flow in such EMRI would produce sufficient dephasing that becomes visible in LISA observation windows. Moreover, the dephasing is very sensitive to the selected model of hydrodynamic disk. One may constrain the disk parameters from this observation.

\section{Discussion and Conclusion}
\label{dissc}

We conclude that there exist observable effects of the disk, though small compared to the leading order amplitude of the GW, still detectable by the proposed detector LISA. The drag effects become significant mainly in the inner region of the disk hence the effect is prominent in the late stage of inspiral and varies considerably with the hydrodynamic model of the accretion flow. The drag effect is more prominent for transonic accretion flow than in other disk models. This could be the reason that in the early studies, the effect of the disk was not considered to be important for the purpose of computing templates for matched filtering techniques. We study the effect of hydrodynamic drag in a  relativistic framework (e.g. \citet{2008PhRvD..77j4027B} ) which offers an accurate measurement of disk-companion interaction than that of the previous Newtonian predictions. Hence the phase acquired by the companion in the real or ideal E/IMRI systems is different and detectable \citep{2019MNRAS.486.2754D,2011PhRvD..84b4032K} which we have obtained here accurately.

As the strength of the hydrodynamic drag and therefore, the emitted GW profile markedly varies with the hydrodynamic models of the accretion flow, one can infer the nature of the accretion disk by observing the GW from EMRI with the disk \citep{10.1007/978-3-319-94607-8_4}. Thus the gravitational wave can serve as a probe to study the properties of accretion disks.

The existence of super Eddington accretion flow will be very helpful to explain the formation of supermassive black holes from the seed stellar mass black holes in the early epoch. In general, there exist several numerical and semi-analytic models that predict the existence of super-Eddington accretion flow \citep{2014ApJ...796..106J,2019ApJ...880...67J,Wang_1999,2013PhRvL.110h1301W}. For such an accretion model radiating efficiency is less than 5-7$\%$ \citep{2014ApJ...796..106J,2019ApJ...880...67J}. Therefore for such low efficiency, a disk with a super-Eddington accretion rate may appear as a sub-Eddington or low luminous object in the optical wavelength. Even when the accretion rate is very high ($>100$ Eddington) the luminosity does not increase proportionately, rather there exists a logarithmic relationship between the accretion rate and the saturated luminosity of this disk \citep{2013PhRvL.110h1301W}. Since GW observation is directly sensitive to the accretion rate, one could be able to measure the accretion rate directly from GW observation, which will be useful to verify the relationship mentioned above. Our accretion model does allow a super Eddington accretion rate. However, this is beyond the scope of our present study and can be done elsewhere.

Moreover, till date, accretion processes are understood only through the EM signals only, and there exist several alternative models of the accretion disks. The sensitivity of GW profiles on the hydrodynamic properties of the disk allows one to test independently the various paradigms of accretion physics  \citep{1996PhRvD..53.2901C}. Thus the simultaneous observations of both GW and EM-waves will enable one to constrain the accretion parameters and to examine the various paradigms of accretion physics. 

With the increased sensitivity of the future GW detectors non-negligible effects on GW, which were once neglected for being small, are now expected to get detected. An ideal E/IMRI study, therefore, needs to be replaced by a real E/IMRI. 
GW is the most efficient probe to understand the physics in the near horizon region as it escapes with impunity from the highly dense matter-rich region almost losing no information which the E.M wave can not. Therefore future scientific studies like testing the paradigms of modified gravity or the existence of exotic stars rely on the GW observations. However, this requires very high precision measurement of GW (e.g. $h \sim 10^{24-26}$) \citep{BOGDANOS2010236}. In such a high level of accuracy, the effect of the disk will certainly be visible and must be eliminated to get the correct result.

\section*{Acknowledgements}
Author SC thanks DST Women Scientist Scheme-A (WOS-A) program for providing fellowship (Feb 2019-Jan 2022) and authors SM and PB acknowledge the associateship programme of IUCAA, PUNE.

\bibliography{gw1-ref}{}
\bibliographystyle{aasjournal}

\end{document}